# Crypto Wash Trading[*]

Lin William Cong       Xi Li       Ke Tang       Yang Yang

First draft: December 2019; current draft: July 2021

## Abstract

We introduce systematic tests exploiting robust statistical and behavioral patterns in trading to detect fake transactions on 29 cryptocurrency exchanges. Regulated exchanges feature patterns consistently observed in financial markets and nature; abnormal first-significant-digit distributions, size rounding, and transaction tail distributions on unregulated exchanges reveal rampant manipulations unlikely driven by strategy or exchange heterogeneity. We quantify the wash trading on each unregulated exchange, which averaged over 70% of the reported volume. We further document how these fabricated volumes (trillions of dollars annually) improve exchange ranking, temporarily distort prices, and relate to exchange characteristics (e.g., age and userbase), market conditions, and regulation.

**Keywords:** Bitcoin; Cryptocurrency; FinTech; Forensic Finance; Fraud Detection; Regulation

**JEL Classification:** G18, G23, G29.

---

[*]The authors are especially grateful to Deeksha Gupta, Kose John, Evgeny Lyandres, and Tao Li for repeated discussions and detailed feedback. We also thank Marlene Amstad, Mykola Babiak, Kevin Dowd, Valeria Ferrar, Itay Goldstein, Angel Hernando-Veciana, Andrew Karolyi, Dongyongp Lee, Jiasun Li, Laura Xiaolei Liu, Roger Loh, Emmanouil Platanakis, Fahad Saleh, Amin Shams, Rajeev Singhal, Baolian Wang, Shang-jin Wei, Wei Xiong, Scott Yonker and seminar and conference participants and reviewers at the Alibaba Group Luohan Academy Webinar, Australasian Banking and Finance Conference, Behavioral Finance/Corporate Finance/Digital Finance (BF/DF/CF) Seminar Group, Cornell University, Cowles Foundation for Research In Economics Conference on the Economics of Cryptocurrencies, 11[th] CSBF Conference (National Taiwan University), Cowles Foundation Economics of Cryptocurrencies (Macroeconomics) Conference, 1st Crypto and Blockchain Economics Research Conference, Durham University Department of Economics and Finance, Econometric Society World Congress (Bocconi University), 2021 Eastern Finanace Association Annual Meeting, IIF International Research Conference & Award Summit, 13th International Risk Management Conference, Inaugural Machine Laywering Conference: "Human Sovereignty and Machine Efficiency in the Law," 18th Paris December Finance Meeting, Paris FinTech and Crypto Webinar, 60th Southwestern Finance Association Meeting, Sun Yat-sen University, 3rd Toronto FinTech Conference, Tsinghua University PBC School of Finance, 3rd UWA Blockchain and Cryptocurrency Conference, 11th Financial Markets and Corporate Governance Conference, European Financial Management Association Annual Meeting 2021, China International Conference in Finance 2021, 17th Asia-Pacific Association of Derivatives Annual Conference, World Finance Conference 2021, Xi'an Jiaotong University, and the Zhongnan University of Economics and Law for helpful comments. This research was funded in part by the Ewing Marion Kauffman Foundation and National Science Foundation of China, and the authors have no affiliation with or research support from any cryptocurrency exchanges. The contents of this publication are solely the responsibility of the authors. Cong (will.cong@cornell.edu) is at the Cornell University Samuel Curtis Johnson Graduate School of Management; Li (xi.li@newcastle.ac.uk) at Newcastle University Business School; Tang (ketang@tsinghua.edu.cn) and Yang (yangyang_tsu@mail.tsinghua.edu.cn) are at the Institute of Economics, School of Social Sciences, Tsinghua University.

# 1 Introduction

The market capitalization of all cryptocurrencies exceeded 1.5 trillion USD in Feb 2021, and the total trading volume is 8.8 trillion USD in the first quarter of 2020 alone (Helms, 2020). Both financial institutions and retail investors have significant exposure to the cryptocurrency industry (Bogart, 2019; FCA, 2019; Fidelity, 2019; Henry, Huynh, and Nicholls, 2019).[1] Meanwhile, crypto exchanges, arguably the most profitable players in the ecosystem, remain mostly unregulated with less than 1% transactions taking place on regulated crypto exchanges in 2019. In the process of vying for dominance in this lightly regulated market, some exchanges are suspected of gaining an advantage in ways ethically and legally questionable (Rodgers (Forbes), 2019; Vigna (WSJ), 2019; BTI, 2019). One form of such market manipulation is Wash trading--- investors simultaneously selling and buying the same financial assets to create artificial activity in the marketplace, which is known to distort price, volume, and volatility, and reduce investors' confidence and participation in financial markets (Aggarwal and Wu, 2006; Cumming, Johan, and Li, 2011; Imisiker and Tas, 2018).

Against such a backdrop, we conduct the first academic study of wash trading and misreporting on cryptocurrency exchanges. By inspecting the distribution of trade size whose first significant digit should follow Benford's law, should exhibit clustering at round numbers, and whose tail distribution is traditionally described by power law (Pareto-Levy law), we find that most unregulated exchanges wash trade (fabricating trades and acting as the counterparty on both sides to inflate volume).[2] We also estimate that unregulated exchanges on average inflate over 70% of the reported volumes in our sample. Furthermore, we provide suggestive evidence that the misreporting (generically referred to as wash trading) improves their ranking and prominence within the industry, relates to short-term price dispersion across exchanges, occurs more on newly established exchanges with smaller userbases, and has implications for the long-term industrial organization, development, and regulations.

---

[1] Surveys reveal that 22% institutional investors have invested in cryptocurrencies (Fidelity, 2019) and by April 2019 9% of adults have owned Bitcoins in particular (Bogart, 2019). In the UK, 25% consumers could identify "cryptocurrency" and 3% had bought them (FCA, 2019). Between 2016 and 2018, Bitcoin ownership increased from 3% to 5% (Henry et al., 2019).
[2] Wash trading is, according to the U.S. Commodity Exchange Act, "Entering into, or purporting to enter into, transactions to give the appearance that purchases and sales have been made, without incurring market risk or changing the trader's market position." Definition of wash trading from US Commodity Exchange Act can be found at *https://www.cftc.gov/ConsumerProtection/EducationCenter/CFTCGlossary/glossary_wxyz.html*

While industry reports in 2018-2019 constitute whistle-blowers, their analyses are often imprecise, ad hoc, unscalable, and non-transparent (Fadilpašić, 2019). It was far from an established fact when we start to publicly circulate of our study. Practitioners were unsure if wash trading only concerns a few specific exchanges with legal cases or was widespread; neither did they know how regulations play a role. Our goal is not to identify a specific wash trade, but to rigorously establish that wash trading is a rampant, industry-wide issue for the cryptocurrency market.[3] We are among the earliest to provide suggestive evidence for the efficacy of regulation in this industry, which has implications for investor protection and financial stability.[4] Our findings also likely have consequences for ongoing lawsuits and empirical research on cryptocurrencies which frequently reference transaction volumes. Finally, they serve as illustrations of the usefulness of statistical and behavioral principles for forensic finance, with regulatory implications for FinTech and beyond.

Wash trading on crypto exchanges warrants our attention for several reasons. First, crypto exchanges play essential roles in the industry (e.g., Amiram, Lyandres, and Rabetti, 2020), providing liquidity and facilitating price discovery just like traditional exchanges. Many crypto exchanges have expanded into upstream (e.g., mining) and downstream (e.g., payment) sectors, consequently wielding great influence as a complex of trading platforms, custodians, banks, and clearinghouses. Naturally, crypto exchanges constitute an anchoring point for understanding the ecosystem from academic, industrial, and regulatory perspectives. Second, because liquidity begets liquidity, crypto exchanges have strong economic incentives to inflate trading volumes to increase brand awareness and ranks on third-party aggregator websites or media (e.g., CoinMarketCap, CoinGecko, Bitcointalk, and Reddit), which in turn increases the exchanges' profits from transaction fees. Third, while wash trading is largely prohibited in most financial markets and developed economies (IOSCO, 2000), cryptocurrencies are particularly prone to wash trading, under limited regulatory oversight.

We collect cryptocurrency transaction information on 29 major exchanges from the proprietary database maintained by TokenInsight (*www.tokeninsight.com*), a data provider who offers consulting, rating, and research reports for the cryptocurrency-related business. TokenInsight chose the 29 exchanges based on their publicity (rank on third-party websites), representativeness, and API compatibility, and the coverage includes well-known exchanges such as Binance, Coinbase, and Huobi, as well as many obscure ones.[5] Our data cover the period from 00:00 July 09th, 2019 (when TokenInsight started to collect transaction information from these exchanges) to 23:59 November 03rd, 2019 (the time we wrote the first draft). Our data also contain variables including aggregate trading volume, reputation metrics, and exchange characteristics such as exchange age.

---

[3] For example, Ontario Securities Commission's recently allegation that Coinsquare's CEO Cole Diamond directed his staff to wash trade, founder Virgile Rostand designed and implemented the codes, and chief compliance officer Felix Mazer failed to take steps he should have taken to stop the actions (Sinclair, 2020). As part of the settlement agreement reached on July 22, 2020, Coinsquare admitted that around 840,000 illicit wash trades were conducted on the platform, amounting to a total value of around 590,000 bitcoins (BTCs).

[4] After we publicly circulated our study in late 2019, several ranking websites changed their matrices from purely volume-based to more sophisticated multi-dimensional ranking models, with at least one website doing so in response to our research finding. Regulators also increased scrutiny on wash trading behavior: Canada-based crypto trading platform Coinsquare has agreed to settle with the Ontario Securities Commission for wash trading charge; Coinbase was fined before IPO for wash trading several years earlier.

We adopt the definition of regulated exchanges from the state of New York, which has one of the earliest regulatory frameworks in the world.[6] For each exchange, we focus on the trading of the four most widely recognized and heavily traded cryptocurrencies against US dollars (USD) — Bitcoin (BTC), Ethereum (ETH), Litecoin (LTC), and Ripple (XRP). We use web traffic ranking as a proxy for brand awareness and reputation to further categorize unregulated exchanges for easy reference: "Tier-1" for exchanges ranking in the top 700 in the finance/investment section of SimilarWeb.com and "Tier-2" for the rest of unregulated exchanges on our data (all ranking outside top 960).

Our first key finding is that wash trading broadly exists on unregulated exchanges but is absent on regulated exchanges. We are fully aware of the challenges of forensic finance and employ multiple approaches that have been successfully applied in numerous fields in sciences and social sciences and are shown to be unlikely affected by dispersed traders' strategies, exchange characteristics, or specificities of the asset class.

To start, we examine the first significant digit for each transaction and check its frequency distribution on each exchange against Benford's law — the well-known statistical benchmark in natural sciences and social sciences and widely used to detect frauds in macroeconomic, accounting and engineering fields (e.g., Durtschi, Hillison, and Pacini, 2004; Li, Cong, and Wang, 2004). We next exploit a classical behavioral regularity in trading: clustering at certain transaction sizes. Round numbers are routinely used as cognitive reference points in individuals' decision-making (e.g., multiples of 10 as cognitive reference points in the decimal system, Rosch, 1975). Rounding is commonly observed in finance (Chen, 2018; Kandel, Sarig, and Wohl, 2001; Kuo, Lin, and Zhao, 2015; Mitchell, 2001), including analysts' forecasts (Clarkson, Nekrasov, Simon, and Tutticci, 2015; Roger, Roger, and Schatt, 2018) or LIBOR submissions (Hernando-Veciana and Tröge, 2020). Most cryptocurrencies are traded at some base units of mental accounts, we thus expect that trades concentrate around multiples of 100, 500, 1000, 5000 and 10000 base units---a natural clustering effect at round sizes. Our third test explores whether the distributions of observed trade size have fat tails characterized by the power law as seen in traditional financial markets and other economic settings (e.g., Gabaix, Gopikrishnan, Plerou, and Stanley, 2003a). We fit a power-law distribution and estimate the exponent parameters in addition to graphically inspecting the tail distributions on a log-log scale. We consistently find anomalous trading patterns only on unregulated exchanges, with Tier-1 exchanges failing more than 20% of the tests and Tier-2 exchanges failing more than 60%. The findings remain robust when under joint hypothesis tests.

---

[5] One is understandably concerned about the potential bias that our data over represent exchanges that wash trade more. However, the data cover a wide range of ranks (1-300 among crypto exchanges) and as we show later, lower-ranked exchanges also have high incentives to wash trade. Moreover, TokenInsight does consider representativeness beyond ranks in choosing the exchanges.

[6] Regulated exchanges are issued BitLicenses and are regulated by the New York State Department of Financial Services. Bitlicence carries some of the most stringent requirements. Our main results are robust to alternative classifications of regulated exchanges. As of June 2020, NYDFS has issued licenses to 25 regulated entities, six of which provide crypto exchange service. They are Itbit, Coinbase, Bitstamp, Bitflyer, Gemini, and Bakkt (futures and options only). Further information can be found at: *https://www.dfs.ny.gov/apps_and_licensing/virtual_currency_businesses/regulated_entities*. (Last accessed: July 3, 2020)

Besides identifying exchanges that wash trade, we quantify the fractions of fake by taking advantage of the rounding regularity. To achieve scale without being easily detected, illicit traders routinely submit machine-generated fake orders and limit the order size (e.g., Vigna and Osipovich, 2018; Rodgers, 2019). Therefore, wash trades primarily generated by automated programs are likely to have low levels of roundness, i.e., a larger effective number of decimals for trades. It is possible that authentic trades are unrounded due to algorithmic trading or other transaction needs. We thus adopt a benchmark ratio (based on calculations from the regulated exchanges) of unrounded trades to authentic trades with round sizes. The extra unrounded trades above the ratio naturally constitute wash trades on unregulated exchanges.

We find that the wash trading volume on average is as high as 77.5% of the total trading volume on the unregulated exchanges, with a median of 79.1%. In particular, wash trades on the twelve Tier-2 exchanges are estimated to be more than 80% of the total trade volume, which is still over 70% after accounting for observable exchange heterogeneity. Combined with the reported volumes in Helms (2020), these estimates translate into wash trading of over 4.5 Trillion USD in spot markets and over 1.5 Trillion USD in derivatives markets in the first quarter of 2020 alone. To mitigate the influence of heterogeneity of traders and algorithmic trading strategies across various exchanges, we validate the roundness-ratio estimation and conduct a number of robustness tests to allay selection concerns.

We then study exchange characteristics that correlate with wash trading and investigate the impact of wash trading on market outcomes such as exchange ranking. In addition, we obtain proprietary data on historical ranking and trading volume information from CoinMarketCap and show that exchange ranking depends on wash trading (70% wash trading of total reported volume moves an exchange's rank up by 46 positions). We find that an exchange's wash trading is positively correlated with its cryptocurrency prices over the short term. We also find that exchanges with longer establishment history and larger userbase wash trade less. Less prominent exchanges, in contrast, have short-term incentives for wash trading without drawing too much attention. Moreover, wash trading is positively predicted by returns and negatively by price volatility.

While current business incentives and ranking systems fuel the rampant wash trading on unregulated exchanges, the regulated exchanges, having committed considerable resources towards compliance and license acquisition and facing severe punishments for market manipulation (Perez, 2015), do little wash trading. We offer a concrete set of tools for exchange regulation and third-party supervision in the crypto market for convincingly exposing wash trading and potentially combating non-compliant exchanges. Admittedly, the tests we introduce are not exhaustive, and wash traders may adjust their strategies in response to these tests. Our tools nevertheless serve as valid detections of wash trading historically and thus make fabrications more difficult and facilitate regulatory resource allocation.

*Literature* — We contribute to recent studies on cryptocurrencies in several ways.[7] Our paper provides the first academic study of crypto wash trading as an industry-wide phenomenon. Existing media evidence is anecdotal and speculative. Industry reports use methods that are not transparent or robust, and they do not typically distinguish regulated from unregulated exchanges. Not to mention that the estimates are often on a small sample, imprecise, and ad hoc.[8] We use rigorous statistical tools and intuitive behavioral benchmarks to establish the existence of wash trading on unregulated exchanges and for various cryptocurrencies. In particular, our tests are not a subset of the tests used by practitioners.

Our paper is most closely related to Amiram, Lyandres, and Rabetti (2021), which builds on our work to offer additional detection tools for wash trading, provide lower bounds using more recent data, and analyze how exchange competition interacts with exchange operations. Aloosh and Li (2021), a subsequent complementary study, verifies our detection methodology by showing individual traders clear their own orders using data leaked from the now-closed Mt. Gox exchange.[9] Another recent study, Le Pennec, Fiedler, and Ante (2021), also adds alternative detection tools utilizing, for example, web traffic or wallet data.

---

[7] Cong, Li, and Wang (2019, 2020), Lyandres, Palazzo, and Rabetti (2020), Howell, Niessner, and Yermack (2020), and Cong and Xiao (2020) provide further institutional background on cryptocurrencies and ICOs; a number of articles discuss the role of crypto-tokens in fundraising and commitment (e.g., Goldstein, Gupta, and Sverchkov, 2020); studies such as Liu and Tsyvinski (2018) and Shams (2020) document empirical patterns in cryptocurrency returns.
[8] Among the earliest whistle-blowers, Bitwise Asset Management presented an industry report to the SEC on March 20, 2019, suggesting potential wash trading on crypto exchanges (Fusaro and Hougan, 2019).

Most of the academic literature on wash trading in traditional markets focuses on investor behavior (e.g., Grinblatt and Keloharju, 2004). We are the first to establish wash trading at the exchange level in the new crypto markets. More broadly, our study belongs to the literature on manipulation and misreporting in finance.[10] Concerning cryptocurrency markets, Foley, Karlsen, and Putniņš (2019) study the illegal usage of cryptocurrencies; Gandal, Hamrick, Moore, and Oberman (2018) and Griffin and Shams (2020) discuss manipulative behavior in Bitcoin and Tether; Li, Shin, and Wang (2020), among others, document pump-and-dump patterns in various cryptocurrencies; most recently, Choi and Jarrow (2020) discuss crypto bubbles caused by speculation or manipulation. These studies do not examine wash trading, which our unique and comprehensive data set allows us to do using robust yet straightforward procedures.

Our study is among the earliest studies on the potential effects of regulation in the cryptocurrency markets, filling in a void in the literature and offering new insights on cryptocurrency regulation. We further speak to the debates on market concentration, collusion, and regulation in the blockchain industry (e.g., Cong and He, 2019; Cong, He, and Li, 2020; Alsabah and Capponi, 2020; Rasu and Saleh, 2020; Lehar and Parlour, 2020; Amiram et al., 2020) by highlighting another detriment of vertical-concentration of the operation scope of crypto exchanges. Related, Irresberg, John, and Saleh (2020) document that only a few blockchains dominate the public blockchain ecosystem.

In terms of methodology, we enrich the use and demonstrate the efficacy of statistical laws and behavioral principles for manipulation detection at scale in accounting and finance. In particular, we are the first to apply Benford's law, trade-size clustering, and power law in FinTech and cryptocurrency studies. Our use of Pareto-Levy distribution (instead of Zipf's law, as seen in Mao, Li, and Fu, 2015 and Prandl et al., 2017) for fraud detection is also novel in social sciences. Finally, our findings imply that researchers and econometricians using reported volumes by exchanges also need to heed the presence of heavy wash trading and test the robustness of their conclusions.

---

[9] The 2019 version of the paper, which was titled "Offchain Bitcoin Trading Puzzle," argues that wash trading on Mt. Gox was limited and could not drive Bitcoin's offchain transactions.
[10] Our paper therefore adds to forensic finance and accounting—the use of economic and financial knowledge to discover or substantiate evidence of criminal wrongdoing that meets standards in a court of law (e.g., Allen and Gale, 1992; Jarrow, 1992; Christie and Schultz, 1994; Ritter, 2008; Zitzewitz 2012).

The paper proceeds as follows. Section 2 describes our data and summary statistics. Section 3 presents the methodologies of wash-trading detection and reports our empirical findings. Section 4 quantifies wash trading and presents an array of tests to validate the methodology and demonstrate the robustness of the results. Section 5 relates wash trading to exchange characteristics, cryptocurrency returns, and exchange ranking, before discussing its implications for regulation and industry practice. Section 6 concludes. Appendix A introduces the development and regulatory status of cryptocurrency exchanges while Appendix B contains the names of the crypto exchanges in our sample. Online appendices provide supplementary evidence and discussion and are available at https://sites.google.com/site/linwilliamcong/CWTOA.pdf.

## 2 Data and Summary Statistics

Our data come from multiple sources. Cryptocurrency transactions are from TokenInsight, which provides ratings and industry reports as an independent third-party. Each transaction is fetched through the exchange's official API (Application Programming Interface) and contains the exchange information, unique transaction ID, timestamp, price, amount of cryptocurrency traded, and trade pair symbol.[11] Our data cover the reported trade history of 29 major exchanges which include all available cryptocurrency trades over the three months from 00:00:00 July 09$^{th}$ to 23:59:59 November 03$^{rd}$, 2019. We then limit the sample to trades of four major cryptocurrencies, Bitcoin (BTC), Ether (ETH), Ripple (XRP), and Litecoin (LTC), representing over 60% of the volume and are available on almost all exchanges. The final sample contains 448,475,535 transactions.

Exchange-related data are collected from both their official websites and various data tracking and analysis platforms. We gather data on exchange ranking, web traffic, etc., from SimilarWeb, Alexa, and CoinMarketCap.[12]

The 29 crypto exchanges in our sample are classified as either regulated or unregulated. The regulation entity of New York State, the New York State Department of Financial Services (NYSDFS), was one of the first agencies to establish regulation over cryptocurrencies and led the world in developing the regulatory framework for the cryptocurrency industry.[13] Hence, we categorize the three exchanges (labeled as R1, R2, and R3) with BitLicense issued by NYSDFS as regulated exchanges because all three operate under the supervision of NYSDFS.[14] BitLicense requires an exchange to build a sophisticated compliance system, an anti-money laundering program, a capital control and custodian system, a record-keeping and customer identity system, an information security team, and a disaster recovery system, as well as to submit necessary documents for routine checks, which cost between 20k to 100k US dollars even for compliant exchanges (Perez, 2015).

---

[11] Since US dollars (USD) are only allowed to exchange in three US regulated exchanges (R1, R2 and R3), digital dollars (e.g. Tether-symbol USDT, which is designed to be pegged to the US dollar) are commonly used as substitutes and widely accepted by the majority of trading platforms, we treat cryptocurrency-USD pairs and cryptocurrency-USDT pairs as being the same.

[12] SimilarWeb and Alexa are online platforms that track and analyze website popularity and provide quarterly rankings by web traffic CoinMarketCap is arguably the most dominant data aggregator and provider in the industry, from which we obtain data on exchange trading volumes and ranks of about 300 exchanges mostly based on daily transaction volumes during the sample period. SimilarWeb ranking is based on a report over the period from Aug 2019 to Oct 2019 *https://www.similarweb.com/*; Alexa historical ranking is accessed through *https://www.alexa.com/siteinfo* on November 15, 2019 and CoinMarketCap ranking is from proprietary data from *https://coinmarketcap.com/*.

The other 26 non-compliant exchanges are classified as unregulated and are further divided into 10 Tier-1 unregulated (labeled as UT1, UT2... UT10) and 16 Tier-2 unregulated exchanges (labeled as U1, U2... U16) based on their web traffic. Web traffic measures reflect an exchange's userbase and reputation and play essential roles regarding customer acquisition and competition. Specifically, Tier-1 unregulated exchanges are the ones in the top 700 of the "SimilarWeb" website traffic ranking of the investment category during the sample period. [15]

Japanese Financial Services Agency (FSA) similarly regulates cryptocurrency exchanges. Subsidiaries of UT5 (Huobi) and UT8 (Okex) are licensed in Japan. From January 10, 2020, crypto exchanges operating in the UK are also required to register with the Financial Conduct Authority (FCA) for anti-money laundering and counter-terrorist financing (AML/CTF) supervisor. In our sample, R2, R3, and UT1 (Binance) have registered with the UK FCA (by September 2020 reference: https://register.fca.org.uk/s/). Our main findings are robust to using these alternative definitions of regulation. For example, UT1, UT5, and UT8 behave in a way more like the regulated exchanges in our baseline definition, than to the average unregulated ones, with only one or two failed tests and compliance with Benford's law for all trading pairs.[16]

---

[13] There is no regulatory framework at the federal level in the United States. Each state is regulating/treating cryptocurrency businesses differently. There are some general requirements based on traditional financial regulations such as compliance AML, KYC, foreign exchange service, money transmitter license, etc. But NY is the only one to introduce this crypto specific license, which is mandatory for exchanges operating in the state and is valid in all other states. Besides, NY is very important in the finance industry because it has always been an important financial hub. Several other countries are actively engaged with crypto businesses, although they have no specific regulations or laws designed for crypto exchanges. For example, Singaporean authority attempts to integrate crypto exchanges into the existing systems by requiring crypto exchanges to comply with the new Payment Services Act (PSA). See Monetary Authority Singapore (*www.mas.gov.sg/regulation/payments/entities-that-have-notified-mas-pursuant-to-the-ps-esp-r*). The Swiss government is actively drafting an Amendment to include Distributed Ledger Technology (a synonym of blockchain technology) into existing Federal Acts (*www.finma.ch/en/authorisation/fintech/*).

[14] Appendix B contains the corresponding names of the exchanges.

[15] The remainder of the unregulated exchanges in our sample all ranked lower than 960. SimilarWeb and Alexa are the two ranking websites based on web traffic. This distinction of tiers does not affect any of our results since they are mostly at the exchange level. The reference to the two tiers simply reflects the differential publicity of the unregulated exchanges and how it correlates with wash trading.

*[Insert Table 1]*

Table 1 summarizes the characteristics of exchanges, including age, trading volume, and ranks from different metrics. Note that age for exchanges refers to the period from their dates of establishment to July 2019. In Table 1, all the regulated exchanges have survived for at least five years to date. However, most unregulated Tier-2 exchanges were launched in 2017 and 2018, while Tier-1 exchanges are generally older. The patterns hint that exchanges benefit from the long-term operation.

Trade volume shows little correlation with our classification of exchanges: Some unregulated exchanges have much larger trading volumes compared with regulated exchanges. For example, U4, an unregulated Tier-2 exchange, has a 50,944 million USD volume while R2's volume is only 15,212 million USD. The trading volume of different unregulated exchanges varies significantly. U9 has only dozens of millions, while a large fraction of unregulated exchanges exceeds tens of billions in the sample.

We find regulated exchanges, especially R1 and R3, fall behind many unregulated Tier-1 exchanges in their ranking based on web traffic. R2 has the highest trading volume among regulated exchanges and a better rank under both ranking algorithms. In terms of CoinMarketCap's ranks based on trading volumes, seven unregulated Tier-2 exchanges rank Top 20 and outperform the majority of unregulated Tier-1 and regulated exchanges. Although trading-volume ranks cannot fully represent the quality and liquidity of exchanges, it is used by most ranking agencies. Thus, cryptocurrency investors are likely to choose an exchange based on these trading-volume based ranks. One would anticipate that unregulated exchanges, especially ones that are launched later, are motivated to engage in wash trading in order to achieve higher rankings and acquire more customers.

Finally, to relate wash trading and crypto exchange ranking, we also acquire proprietary, high-frequency data on exchange ranks and reported trading volumes from coinmarketcap.com. The platform started its business by providing crypto market capitalizations, pricing, and other information on all kinds of cryptocurrencies. Growing together with the industry, the company has become a top data provider and ranking agency in the industry. As of June 12, 2020, it serves 4.2 million unique visitors around the globe with 32.6 million visits per month (SimilarWeb.com), dominating its kind with a valuation in the Binance acquisition proposal (not publicly disclosed) in March 2020 believed to be 400 million USD (Bambrough, 2020). Currently, this "Crypto Standard and Poor's" declares itself as accurate and neutral. However, given their influence and vital function, these third-party rating agencies are likely to face more regulation just like credit rating agencies in traditional financial markets.

---

[16] That said, their trade-size roundness differs from the regulated exchanges in our baseline categorization. While they are still distinct from most other unregulated exchanges, they do have an estimate of more than 50% of the volume on average being wash trades. This could be reflections of the more stringent regulatory standard of NY Bitlicence, but could also be attributed to the fact that UT5 and UT8 only have subsidiaries regulated in Japan and FCA did not mandate the regulation of UT1 during our sample period.

## 3 Empirical Evidence of Wash Trading

We present empirical evidence of crypto wash trading entailing four major trading pairs (BTC/USD, ETH/USD, LTC/USD, and XRP/USD).[17] Specifically, we examine the properties of trade sizes on each exchange and test them against three well-established statistical and behavioral benchmarks. The multitude of statistical tests when reporting at the exchange level demonstrates the presence of wash trading on unregulated exchanges in a robust manner. Because they are based on fundamental behavioral and statistical principles, they are the least prone to the influence of heterogeneous (but authentic) trading specific to individual traders and exchanges, which we further control for when quantifying the extent of wash trading in the next section.

### 3.1 Distribution of First Significant Digits

We investigate whether the first-significant-digit distribution of transactions (denominated in the cryptocurrencies in question) on each exchange conforms to the pattern implied by Benford's law. Inconsistency with Benford's law suggests potential manipulations.

**3.1.1 Benford's Law**

Benford's law describes the distribution of first significant digits in various naturally generated data sets and derives from the intuition that many systems follow multiplicative processes (e.g., Li, Cong, and Wang, 2004).[18] According to Benford (1938):

---

[17] Our choice of trading pairs is motivated by brevity and dominance. LTC/USD data is not available in unregulated exchange UT7, U1, U6, and U9. XRP/USD data is not available in regulated exchange R3 and unregulated exchanges U1 and U6. Trading pairs involving other cryptocurrencies exhibit similar patterns.

[18] Benford's law, also known as Newcomb–Benford law, was first proposed by the American astronomer Simon Newcomb in 1881 after observing the degree of abrasion in different parts of books in a library. Though initially unnoticed, the

$$\text{Prob}(N \text{ is the first significant digit}) = \log_{10}(1 + N^{-1}), \, N \in \{1,2,3,4,5,6,7,8,9\}. \quad (1)$$

The probability of 1 being the first significant digit is 30.10%. Digits 2 and 3 have probabilities of 17.60% and 12.50%, respectively. The probabilities of the rest (9.7%, 7.9%, 6.7%, 5.8%, 5.1%, and 4.6%, respectively) being the first significant digits decrease as the digit increases.

Naturally, Benford's law holds in data sets randomly and independently generated from one distribution or mixed random sampling from various distributions. Apart from natural or sequential data (e.g., mobile numbers), deterministic samples with exponential growth or decay also follow Benford's law or its variants when numbers are expressed in different bases. Benford's law has been effectively applied to test the reliability of data and detect manipulation or anomalous patterns in a wide array of data sets.[19]

**3.1.2 Detecting Violations of Benford's Law**

We check whether the leading digits of trade sizes follow Benford's law (as shown in Equation 1) on the 29 exchanges. Figure 1 illustrates the first-significant-digit distribution for four cryptocurrencies with one regulated exchange and four unregulated exchanges. The five exchanges are the ones that fail the most tests in their categories and are consistently chosen throughout the paper for concise illustration. The distributions for the rest of exchanges exhibit similar patterns and are shown in Online Appendix A. Bars show the fraction of transactions in which the trade size has integer *i* as the first significant digit. Dots represent the frequency distribution implied by Benford's law.

*[Insert Figure 1]*

For R2, 32.75% of BTC trades and 32.73% of ETH trades have "1" as the leading digit, consistent with the benchmark frequency of 30.10% in Benford's law. Unregulated exchanges such as U8 and U9 clearly violate Benford's law, with some first significant digits occupying a disproportionally large fraction. In general, first-significant-digit distributions of all regulated exchanges comply with

---

proposed law was rediscovered and elaborated in detail by the American physicist Frank Benford (1938). It is applicable in trading (and has been empirically verified in various asset markets) because reinvesting excess returns and reducing budget after losses makes the budget process a multiplicative process.

[19] Prior literature provides statistical evidence for Benford's law (e.g., Hill, 1995, 1998; Pinkham, 1961). Li, Cong, and Wang (2004) provide an overview. Sambridge, Tkalčić, and Jackson (2010) find that Benford's law holds for 15 sets of modern observations drawn from the fields of physics, astronomy, geophysics, chemistry, engineering, and mathematics. In economics, Benford's law is introduced for fraud detection in tax payments, accounting, macroeconomics, hospitality management, international trade, and finance (Durtschi et al., 2004; Nigrini, 1996; Günnel and Tödter, 2009; Gonzalez-Garcia, 2009; Liu and Moulton, 2018; Liu, Sheng, and Wang (2020); Chakrabarty et al., 2020). To see why Benford's law intuitively applies to trading, imagine a portfolio manager trading cryptocurrencies only, with all returns reinvested. Then the size of the portfolio and thus the transactions roughly follow a stochastic multiplicative process. For example, one might be trading 15-19 bitcoins initially, but as the crypto wealth grows, the transactions could become 20+ or 30+ bitcoins. If the portfolio incur losses, the transactions might go down to 10-14 bitcoins. Benford's law would capture this pattern of having first significant digits being 1 more frequently than being other digits.

Benford's law regardless of the type of cryptocurrency. For unregulated exchanges, including Tier-1 and Tier-2, half of them exhibit apparent discrepancies with Benford's Law in at least one type of cryptocurrency. Disconformity with Benford's Law is observed on nine unregulated Tier-2 exchanges, among which seven violate the law in at least two cryptocurrencies.

*[Insert Table 2]*

We employ the Pearson's Chi-squared test to quantitatively assess whether first-significant-digit distributions conform with Benford's law (see Table 2). Trades of regulated exchanges follow Benford's law, so do those on most of the unregulated Tier-1 exchanges. However, patterns for UT3 are inconsistent with Benford's law in BTC and XRP trades, with a significance level of 1%. Moreover, five Tier-2 exchanges (U5, U7, U8, U9, and U14) have significant divergence from Benford's law in most cryptocurrencies. Other unregulated exchanges show sizable differences in several cryptocurrencies. For example, UT7 violates Benford's law in BTC at a 5% level; U2 and U10 fail in BTC and XRP at a 1% confidence level, respectively; U2 and U3 fail at a 5% confidence level in ETH.

Overall, all regulated exchanges show consistency with Benford's law; 20% of unregulated Tier-1 exchanges violate Benford's law in at least one cryptocurrency, at a 5% significance level; 50% of Tier-2 exchanges fail to follow Benford's law in at least one cryptocurrency.

### 3.2  Trade Size Clustering

As a second test, we investigate whether the trades on crypto exchanges also feature clustering—traders' tendencies to use round trade sizes and round prices—, the classical behavioral regularity commonly observed in financial markets.[20] Clustering occurs because authentic traders tend to use round numbers as cognitive reference points (Rosch, 1975) to simplify and save effort in the decision-making and evaluation process (Ikenberry and Weston, 2008; Kuo et al., 2015; Lacetera, Pope, and Sydnor, 2012). Therefore, the cognitive reference of round numbers sets authentic trades apart from robot trades (Mahmoodzadeh and Gençay, 2017; O'Hara, Yao, and Ye, 2014). Because wash traders use machine-based automated trading programs to save manpower, especially when fake orders feature small trade sizes but large total amounts (Vigna and Osipovich, 2018; Rodgers, 2019), wash trading naturally reduces the proportion of authentic volume, and thus clustering.

---

[20] For instance, Alexander and Peterson (2007) show that in the New York Stock Exchange (NYSE) and Nasdaq, higher proportions of trades occur at round sizes that are multiples of 500, 1000 or 5000 shares compared to other sizes. Verousis and ap Gwilym (2013) find trade size clusters at multiples of 500 shares on the London Stock Exchange. Mahmoodzadeh and Gençay (2017) document the human's preference for round prices after exchanges change their decimal price systems. Clustering is also observed in foreign exchanges (Moulton, 2005), derivative markets (ap Gwilym and Meng, 2010), and the U.S. equity market (Ikenberry and Weston, 2008).

Because most cryptocurrencies can be traded in fractions, and some currencies have larger unit values (especially BTC), we set in the remainder of the paper the smallest unit (base unit) to be one unit in a certain decimal place valued in the neighborhood of one US dollar. For instance, with the price of Bitcoin varying around $8000-$10000 in our sample period, most BTC-USD orders are below 1 BTC. Therefore, round numbers in traditional financial markets such as 100, 1000, or 10000 are too big for individual traders. Because the value of $10^{-4}$ BTC is in the order of magnitude of one US Dollar, it is natural to consider $10^{-4}$ BTC as the base unit in this study. Similarly, the base units of ETH, LTC, and XRP are 0.001 ETH, 0.01 LTC, and 1 XRP, respectively. We now examine whether trade-size clustering appears at multiples of 100 base units for each cryptocurrency.[21]

### 3.2.1 Histograms of Trade Size

Figure 2 depicts trade size distributions of representative exchanges in two observation ranges for BTC, ETH, LTC, and XRP, highlighting the clustering effect at the round sizes.[22] Online Appendix B displays the histograms of the remaining exchanges. Panel R, Panel UT and Panel U depict the trade-size distribution for regulated exchanges, unregulated Tier-1 exchanges, and unregulated Tier-2 exchanges, respectively. Note that the Y-axis represents the probability that transactions fall into each interval, shown on a log scale.

*[Insert Figure 2]*

Firstly, three regulated exchanges (R2 in Figure 2; R1 and R3 in Online Appendix B) display a downward sloping curve with prominent peaks at multiples of 5000 base units in the range of 0-10 BTC (e.g., 0.5 BTC, 1 BTC, 1.5BTC, 2BTC, etc.). Similar patterns also appear in distributions of ETH, LTC, and XRP. The findings suggest the presence of trade size clustering on regulated crypto exchanges. This finding is consistent with the trade pattern in regulated financial markets, which display a downward trend because large orders are less frequently placed and executed, as well as a trade size clustering effect (e.g., Alexander and Peterson, 2007; ap Gwilym and Meng, 2010; Mahmoodzadeh and Gençay, 2017; Verousis and ap Gwilym, 2013). Similar to participants in traditional markets, cryptocurrency investors exhibit preferences for round trade size.

---

[21] We focus on clustering in terms of round numbers in the number of tokens instead of dollar amounts because our data contains the number of tokens traded and its product with token price is typically not equal to the actual dollar amount traders use in their orders due to exchange fees. For a few exchanges that we can obtain the time series of fees, we find our results to be robust to the alternative specification using dollar amounts.

[22] The observation ranges include 0-1 BTC, 0-10 BTC, 0-10 ETH, 0-100 ETH, 0-100 LTC, 0-1000 LTC, 0-10000 XPR, and 0-100000 XPR.

Taking Bitcoin as an example, UT6 in Figure 2 does not show clear clustering patterns. Besides, most trades of UT6 are concentrated at small sizes and display an anomalous drop in frequency, especially in LTC and XRP trades. Moreover, clustering patterns for different assets vary across crypto exchanges and have shown no overall pattern. [23]

On unregulated Tier-2 exchanges, we observe less apparent clustering at round sizes. Moreover, trade patterns vary dramatically and are distinguishable from the typical downward distribution. For instance, trade frequency on U8 does not monotonically change with the increase in trade size in all cryptocurrency trades when zooming out to larger ranges. Similar issues are observed on other exchanges (see Online Appendix B, e.g., U5, U7, and U15 in BTC trades; U3, U7, U11, and U15 in ETH trades). Additionally, on U8, gaps are observed in the histograms of 0-100 ETH, 0-1000 LTC, and 0-100000 XRP trades. Similarly, transactions on U9 are absent in irregular intervals of trade size, and gaps erratically appear in the range of 0.3-1 BTC, 5.5-9.5 ETH, and 2500-5500 XRP. When zooming out to larger trade-size ranges, trade patterns of U9 exhibit a cliff pattern with a steep decline in all cryptocurrencies. Visually, U14 shows scarce peaks at round sizes of all cryptocurrency trades. A uniform distribution is observed in LTC and XRP, as well as large observation ranges of BTC and ETH.[24] The finding indicates that investors trade with approximately equal frequency at different trade sizes, which is against the behavioral regularity in financial markets.

**3.2.2 Statistical Tests for Clustering**

To quantify the effect of trade-size clustering, we conduct the Student's t-test for each crypto exchange by comparing trade frequencies at round trade sizes with the highest frequency of nearby unrounded trades. For each trading pair, we set up two sets of observation windows: windows centered on multiples of 100 units (100X) with a radius of 50 units (100X-50, 100X+50), and windows centered on multiples of 500 units (500Y) with a radius of 100 units (500Y-100, 500Y+100). Trade frequency is calculated as the number of trades with size $i$ over total trade numbers in the observation window. For example, Figure 3 shows that in BTC trades on R1, the observation window around 200 units (0.02 BTC) ranges from 150 units (0.015 BTC) to 250 units (0.025 BTC). Trades at 0.02 BTC constitute 16.42% of total trades in the 0.015-0.025 BTC range, while the highest trade frequency of unrounded trades is only 2.54% in the observation range. The apparent difference indicates that trades with 0.015-0.025 BTC cluster at 0.02 BTC (200 base units).

---

[23] For some Tier-1 exchanges, clustering is less apparent in the trades of XRP than other cryptocurrencies (see Panel UT2, UT4, and UT5 of Online Appendix B).
[24] Furthermore, at least six Tier-2 exchanges display uniform patterns in cryptocurrency trades (e.g., U1, U2, U3, U6, U10, U11, and U12 in Online Appendix B).

*[Insert Figure 3 and Table 3]*

Table 3 presents the *t*-test results for size clustering on regulated exchanges (Panel A), unregulated Tier-1 (Panel B), and Tier-2 exchanges (Panel C). As expected, on all three regulated exchanges (Panel A in Table 3), trade frequency at round sizes is higher than unrounded ones by a large margin regardless of cryptocurrencies and observation ranges, consistent with our findings in Figure 2. Additionally, size clustering is more evident at multiples of 500 units in terms of difference and *t*-statistics since 5 is at a higher level of roundness than 1. For example, for BTC trades on exchange R1, the difference in frequency is 9.1% in trade size of multiples of 100 units (e.g., 0.01 BTC, 0.02 BTC, and 0.03 BTC) while the difference is 20.3% at the size which is the common multiples of 500 units (e.g., 0.05 BTC, 0.01 BTC, 0.015 BTC). The results are consistent with the rounding behavior.

Similar to regulated exchanges, three unregulated Tier-1 exchanges (UT3, UT7, and UT9) show positive and significant differences at a 1% level in trades of all available cryptocurrencies (except for XRP on UT9, which is significant at 5%). Trade clustering appears more frequently at multiples of 500 units as well: for example, six Tier-1 exchanges (UT1, UT3, UT5, UT7, UT8, and UT9) exhibit noticeable clustering effects at multiples of 500 units for all four cryptocurrencies. However, UT6 and UT10 show insignificant differences in frequencies between round and unrounded trades.

In contrast, clustering at round sizes is largely absent on unregulated Tier-2 exchanges. Half exchanges exhibit no sign of clustering for all cryptocurrencies in both observation windows (100X; 500X). Except for U13, all Tier-2 exchanges have no clustering in at least one cryptocurrency. Besides, on some exchanges, trade clustering becomes less obvious at a higher level of roundness (multiples of 500 units). For example, on U3 and U5, frequencies at multiples of 100 units are higher (significantly at a 1% level), but clusters at multiples of 500 units are not significant.

We also regress the (logit) percentage of trades at a certain size on various dummy variables which are set to one at round sizes. The results (shown in Online Appendix C) are consistent with the tests in this section.

In sum, we document that regulated exchanges display an evident clustering effect in trade size, whereas unregulated Tier-1 and Tier-2 exchanges contain little clustering, with 30% and 50% exchanges displaying no trade-size clustering in all cryptocurrencies, respectively. Note that clustering is about rounding off the last non-trivial digits and affects little the distribution of the first significant digits. To the extent that this is a concern, one can use variants of Benford's law with the first several significant digits for robustness.

### 3.3 Tail Distribution

In this section, we examine the tails of trade-size distributions on each crypto exchange. By fitting the tails with power-law distributions, which adequately describes patterns in traditional financial markets, we can detect anomalous behavior of reported cryptocurrency trades.

**3.3.1 Power-law Distribution as a Statistical and Behavioral Benchmark**

In economics and finance, power law captures the "fat tails" of many distributions, including the Pareto distribution of income (Pareto, 1896), the distribution of stock returns (Gopikrishnan et al., 1999), trade size (Gopikrishnan et al., 2000), and share volume (Plerou et al., 2000; Plerou and Stanley, 2007), fluctuations in foreign exchange markets (Da Silva, Matsushita, Gleria, and Figueiredo, 2007; Ohnishi et al., 2008; Vandewalle, Ausloos, and Boveroux, 1997), and cryptocurrency transactions (Li et al., 2019; Schnaubelt et al., 2019). Gabaix (2016) provides an overview.

Mathematically, the power-law distribution has a cumulative density function (CDF) that follows the form

$$P(X > x) \sim x^{-\alpha} \tag{2}$$

where $\alpha$ is known as the power-law exponent or the tail exponent. When using the probability density function (PDF), the relevant parameter is $\alpha_{pdf} = \alpha + 1$.

One explanation for power-law tails in the empirical data is the trading behavior of large investors, who try to avoid large price impact in the markets (Gabaix, Gopikrishnan, Plerou, and Stanley, 2003a). Other studies attribute the emergence of power-law to the investors' limited information on the value of assets (Kostanjčar and Jeren, 2013; Nirei, Stachurski, and Watanabe, 2018) and herding (Nirei et al., 2018). In the crypto market, large participants (e.g., institutional investors or large retail investors) have increasingly participated in cryptocurrency trading. Investors generally have asymmetric information on the value of cryptocurrency. For all these reasons, transaction sizes are highly likely to conform to the power law.

### 3.3.2 Power Law and Tail Exponents

To examine trade size distribution tails, we used two widely adopted techniques: The first one is to take the logarithm of the empirical probability density function and fit the log-log data to power-law distribution by Ordinary Least Square (OLS). The second one is to apply the Maximum Likelihood Estimation approach (MLE) and use the Hill estimator $\hat{\alpha}_{Hill}$ for the data fitting. Hill estimator is asymptotically normal and calculated as follows (Clauset, Shalizi, and Newman, 2009; Hill, 1975):

$$\hat{\alpha}_{Hill} = 1 + n \left( \sum_{i=1}^{n} ln \frac{x_i}{x_{min}} \right)^{-1} \qquad (3)$$

where $n$ is the number of observations and $x_{min}$ is the cut-off threshold. The distribution yields to power-law after $x_{min}$. The cut-off $x_{min}$, which signifies the start of the tails, is set as the top 10% of the largest trades during the sampling period. In this study, trade size distributions are constructed for empirical probability density functions, and the python package 'powerlaw' (Alstott, Bullmore, and Plenz, 2014) is applied to fit the data and calculate the exponent.

Gabaix, Gopikrishnan, Plerou, and Stanley (2003b) show that stock trade size follows a half cubic law ($\alpha = 1.5$) both theoretically and empirically. Various studies on trading volumes or sizes have shown that the vast majority of tail exponents lie in the Pareto–Lévy regime ($1 < \alpha < 2$) for traditional financial assets and bitcoins (Li et al., 2019; Schnaubelt et al., 2019).[25] We thus check whether the values of exponent α in the fitted results fall within the Pareto–Lévy range.

Table 4 presents the results from OLS and MLE fittings for four cryptocurrency trades. We can visually inspect the goodness of fit and identify whether crypto exchanges display a power-law tail in trade size distribution, shown in Figure 4.

*[Insert Table 4 and Figure 4]*

As expected, both scaling estimators $\hat{\alpha}_{OLS}$ and $\hat{\alpha}_{Hill}$ lie in the Pareto–Lévy regime on regulated exchanges and suggest a stable power-law decay in all cryptocurrency trades. Similar patterns are observed on half of the unregulated Tier-1 exchanges. In contrast, estimators of two Tier-1 exchanges (UT4 and UT6) do not fall into the Pareto–Lévy range for four cryptocurrencies and suggest inconsistency with power-law exponents for trade size in traditional markets. Besides, tail exponents for UT7, UT8, and UT10 are outside the range of 1 to 2 in one cryptocurrency.

---

[25] Gopikrishnan et al. (2000) find that the power law exponent of trade volume is around 1.5 in US equity market. Plerou and Stanley (2007) investigate trades in New York Stock Exchange, London Stock Exchange and Paris Bourse and show that trade size in all three markets display power law decay with exponent in the range from 1 to 2. Moreover, value of exponents is not affected by industry and market capitalization. Note that Mandelbrot (1960) propose that income follows the stable "Pareto–Lévy" distributions with $1 < \alpha < 2$.

On unregulated Tier-2 exchanges, only three exchanges show estimated exponents within the Pareto–Lévy range, whereas 62.5% show statistical evidence in disconformity to parameters of empirical regularity in four cryptocurrencies. For the rest, the estimated exponents of U12 follow Pareto–Lévy range in LTC and ETH trades, while U14 and U16 show a similar fashion in LTC and ETH trades, respectively.

Figure 4 displays the probability density for trade size and the fitted power-law distributions on log-log plots, with one regulated and four unregulated exchanges as representatives for brevity. Online Appendix D contains figures of the rest.

As in mainstream financial markets, transactions from regulated exchanges display a downward linear trend in the log-log plots and appear visually fitting the power-law distribution. For instance, in Panel R2 of Figure 4, empirical data points fall around the fitted lines without obvious outliners, implying that trades in regulated exchange generally follow the power law in all four listed cryptocurrencies. In general, the OLS line fits equally in the whole range, while MLE estimation weighs more at the start of the tail, where the probability value is higher. Consistent with regulated exchanges, 90% of unregulated Tier-1 exchanges resemble power-law tails in trade size distributions. Straight lines estimated by OLS and MLE are roughly fitted to the data. Conversely, UT6 (shown in Figure 4) shows a curvy shape in tails and fails to show the power-law distribution in the trade size.

On unregulated Tier-2 exchanges, tail distributions vary differently and display irregular patterns across exchanges and cryptocurrencies. Four Tier-2 exchanges (U6; U13; U15; U16) show a linear decrease in the tail zones and comply with the power-law tail. U9 (shown in Figure 4) displays a good linear fit but shows inconsistency with the MLE fitted line. On U8, data points disperse in the tails of BTC, ETH, and LTC trades; additionally, a curvy shape is observed on the logarithm scale in BTC and XRP trades. In BTC trades of U14, the tail appears to be level with some outliers far from the line. ETH, LTC, and XRP trades of U14 show a step-like decay.

Combing the results above, regulated exchanges behave as the power law predicts, with estimators consistent with Pareto–Lévy exponents in mainstream financial markets. 50% of Tier-1 exchanges display a power-law tail with exponents characterized by the Pareto–Lévy regime in all cryptocurrencies. 75% of unregulated Tier-2 exchanges fail to follow the Pareto–Lévy power law that is commonly observed in financial markets.

## 3.4 Selection Concern, Multi-hypothesis Testing, and Conclusive Evidence

In our discussion thus far, three independent statistical analyses are conducted for each cryptocurrency of each crypto exchange, including the Chi-squared test for Benford's Law distribution, *t*-test for trade-size clustering, and linear fit for power law.[26] The results are consistent for each category (regulated, unregulated tier-1, and unregulated tier-2) and for the majority of exchanges. Overall, more than half of the unregulated exchanges fail at least half of all tests at the 5% significance level. Except for U13, Tier-2 exchanges fail at least 30% of the tests, with ten exchanges failing more than 65% of all the tests. At the cryptocurrency level, unregulated exchanges as a whole fail more than 40% of the tests for each cryptocurrency. In contrast, regulated exchanges pass all the tests.

Because the multiple statistical tests may increase the possibility of Type I error and raise the concern of *p*-hacking, we perform a multiple (global) hypothesis test on the null hypothesis that trade patterns of crypto exchanges are consistent with universal laws or patterns in traditional financial markets, using Fisher's method for each exchange-currency pair. In Fisher's method, *p*-values from individual tests were combined into a statistic ($\chi^2$) using the formula below:

$$\chi^2_{2n} = -2 \times \sum_1^n \log p_i \quad (4)$$

in which n is the number of independent statistical tests and $p_i$ is the individual *p*-value from the test *i*. Note that the critical value for $\chi^2_6$ at a 5% significant level is 12.592. Larger than that, the null hypothesis will be rejected.

[Insert Table 5]

The results from the multiple hypothesis tests (summarized in Table 5 with more details in Online Appendix E) are consistent with our findings in previous subsections. Trade patterns of all regulated exchanges show insignificant differences from those of traditional financial markets. Tier-1 unregulated exchanges have lower proportions in rejecting null hypotheses than Tier-2 ones in all cryptocurrencies. 75% of the Tier-2 unregulated exchanges fail to follow the universal law or trade patterns of traditional financial markets. In addition, BTC has the highest failure rates, followed by XRP. Furthermore, more unregulated exchanges fail the joint tests than individual tests in all cryptocurrency pairs. Some fraudulent exchanges may "luckily" display similar trade distribution as traditional markets in certain aspects but fail to follow all regularities, therefore leading to higher failed percentages in multiple hypothesis tests.

---

[26] Except for R3, UT7, U1, U6, and U9, 24 crypto exchanges contain the full set of four trading pairs.

One might be concerned that traders and algorithms are special on unregulated exchanges and traders select unregulated exchanges for a reason.[27] First, there is no evidence that the trading strategies or the extent of algorithmic trading are different across various exchanges. On the contrary, trading algorithms are often documented to be close to exchange-agnostic (Alameda, 2019). According to CoinMarketCap, the distribution of institutional investors is primarily correlated with the exchange volume than its regulatory status. We also find no significant difference regarding the volume and distribution of transactions on regulated exchanges compared to unregulated exchanges around the time they became regulated. For example, Coinbase received Bitlicense in 2017. But there is no exodus of traders. If anything, its volume grew significantly.

While these may not completely rule out traders' self-selection into trading on regulated versus unregulated exchanges, we appeal to the power of Benford's law and power law to allay our concern. If institutional investors or algorithmic traders systematically prefer regulated or unregulated exchanges, it would bias our findings towards seeing different tail distributions and less rounding on those exchanges. However, power-law distributions for the tail can have different cut-offs, so having disproportionally large trades is unlikely to affect the general shape of the tail (it could affect the exponent parameter). Moreover, Benford's law entails the first significant digits and is based on multiplicative growth. Whether one transact large amounts or use algorithms should not affect the first significant digits because Benford's law is robust to a change of accounting unit and rounding behavior. Therefore, the three tests really complement one another to help identify wash trading.

In conclusion, this section establishes abnormal trading patterns on unregulated exchanges while suggesting the absence of wash trading on regulated crypto exchanges.

## 4 Quantifying Wash Trading

Given the rampant phenomenon of wash trading across unregulated exchanges involving various cryptocurrencies, we now quantify the extent of wash trading by directly estimating wash trading volume. We also conduct several robustness and validation tests for our estimator and provide alternative metrics such as "certainty of wash trading."

---

[27] Our detection could potentially depend on the volatility of the cryptocurrencies. Our observation window is unfortunately too short for us to detect any significant patterns regarding how volatility affects whether an exchange has wash trading on that cryptocurrency.

## 4.1 Trade-size Roundness and Benchmark Roundness Ratio

Authentic human trades tend to have round sizes. In contrast, unrounded trades typically relate to programmed trading for various purposes such as market marking, high-frequency arbitrage, and in particular, wash trading, which is highly likely to be conducted using automated programs or bots considering the efficiency and quantity of trade orders required. Strong evidence suggests that most wash trading is done by bots, which can be easily added layers in the trading structure scripted by simple Python programs (e.g., Vigna and Osipovich, 2018; Rodgers, 2019).[28] Therefore, round/unrounded trades can be used as a reasonable proxy for authentic orders/fake trades. The roundness of trade size is consistent with the clustering analysis of trade sizes in Section 3.2.

To start, we show that levels of roundness for trade sizes differ across unregulated exchanges and regulated ones. The level of roundness is a qualitative parameter describing the decimal or integer places of the last non-zero digit. For instance, 1.01 BTC has a higher level of roundness than 2.123 BTC; 100 ETH has a higher level of roundness than 1234 ETH.[29] Authentic trades should display a higher level of roundness in size than artificial ones. We thus expect regulated exchanges to present a higher level of roundness in trade sizes compared with unregulated exchanges if we are to use them as benchmarks. For each crypto exchange, we analyze the trade-size distribution over levels of roundness (ten thousands, thousands, hundreds, tens, ones, tenths, hundredths, etc. base units). We compare the distributions for the level of roundness on regulated and unregulated exchanges.

*[Insert Table 6]*

Table 6 shows the Chi-squared tests of the comparison for four cryptocurrencies. All Tier-1 exchanges have significantly large Chi-squared statistics in at least one cryptocurrency. As for unregulated Tier-2 exchanges, except for U7 in BTC trades, all trades show completely different roundness distributions from regulated exchanges with a 1% significance level for nearly all cryptocurrencies. The finding shows that unregulated exchanges, especially unregulated Tier-2 exchanges, have a lower level of roundness in trade size relative to the regulated exchanges.

---

[28] There is no need to explore darknet marketplaces or shady hacking forums or to buy black hat services. One of the bot tools, "Ping-Pong," allows executing simultaneous buy and sell orders to the users themselves, creating a mirage of active trading for particular cryptocurrencies.
[29] For 1.01 BTC, the place value of last non-zero digit (1) is hundredths, while the place value of last non-zero digit (3) is thousandths in 2.123 BTC. In 100 ETH, the place value of last non-zero digit (1) is hundreds while the place value of last non-zero digit (4) is ones in 1234 ETH.

Assuming that the computer-based legitimate (non-wash) trades on unregulated exchanges have the same sensitivity to the authentic trading strategies and exchange characteristics as those on regulated exchanges, we can estimate the legitimate amount of unrounded trades for unregulated exchanges. The difference between the observed unrounded and legitimate trading volume is then a reasonable proxy for the wash-trading volume. Since it is rarely the case that one can directly label wash trades at an exchange without confessions by or detailed information of the traders, our method provides a general way of estimating systematic wash trading that can be time-varying, therefore serving as a first-order benchmark.

From our earlier analysis, we do not detect systematic wash trading on regulated exchanges. This is further corroborated by the fact that round trades constitute around 30% of total trades on regulated crypto exchanges, which is consistent with patterns in the U.S. equity markets that are approximately free of wash trading due to regulation (Gomber, Gsell, Pujol, and Wranik, 2009; Tabb, Iati, and Sussman, 2009). We also carry out a "cross-validation" test. We use any two regulated exchanges as the no-wash-trading benchmark, to estimate the wash trading amount on the remaining regulated exchange. We found the wash trades estimated on average constitute less than 5% of the reported volumes, indicating the absence of clear evidence for wash trading.

## 4.2  Estimated Volume of Wash Trades

We estimate the volume of wash trades by calculating the abnormal proportion of unrounded trades for various exchanges. Specifically, we categorize trading volumes into round and unrounded ones by checking if the last non-zero digit of a certain trade size is less than 100 basis units or not. We then perform a pooled regression to estimate the ratio of (log) unrounded volume to (log) round volume for all regulated exchanges with a weekly frequency:

$$\ln(V_{Unrounded_{it}}) = \alpha + \beta * \ln(V_{Round_{it}}) + \gamma * X_{it} + \epsilon_{it}, \qquad (5)$$

where $V_{Unrounded_{it}}$ and $V_{Round_{it}}$ are unrounded and round trading volumes of regulated exchange *i* at week *t* respectively. In the baseline, we exclude exchange-level controls by setting $X_{it}$ to zero. To mitigate the concern that heterogeneous authentic algorithmic trading on various exchanges drives the estimates, we include a vector of exchange characteristics, $X_{it}$ including age, rank, CoinMarketCap web traffic percentage, and unique visitors, in an alternative specification. We employ the parameters in (5) to calculate the legitimate (non-wash) unrounded trades of unregulated exchanges using their corresponding round trades. Wash trade volumes are thus calculated as the non-negative amount by which the total unrounded trades exceed legitimate unrounded trades.

[Insert Table 7]

Table 7 presents the simple averaged and volume-weighted wash trading percentage for each exchange category, as well as the exchange-level wash trading percentage by four cryptocurrency pairs. The results using models with or without controls are similar. Because some exchanges are missing data on the control variables and the residual standard errors in the model without controls are comparable to the ones with controls (so out-of-sample predictability are comparable), for later analysis on price impacts, ranking, etc., we only report the results using estimates from the model without controls for simplicity. Standard deviations of wash trading volumes from bootstrapping the sample 1000 times are also included in the table.

On average, wash trades account for over 70% of total trading volume on each unregulated exchange, and about 61% even after controlling for exchange characteristics. Wash trades are above 53.4% for Tier-1 and 81.7% for Tier-2 exchanges. Because the four cryptocurrencies we look at dominate the transaction volumes on all the exchanges, the numbers are reasonable estimates even if one includes all cryptocurrencies. It is also worth noting that for all unregulated exchanges, an estimate of 77.5% of the total reported volume appears to be wash trades. Our estimates are in the same order of magnitudes as the estimates from the Wall Street Journal and industry reports (Rodgers, 2019; BTI, 2019), which are in the range of 67% to 99%. For example, the BTI Summary of Market Surveillance report found 17 of the CoinMarketCap top 25 exchanges to contain over 99% fake volumes, as of April 2019. Our estimates are slightly lower because exchanges could have reacted since those earlier estimates were released. So the usual Lucas critique applies. [30]

### 4.3   Further Validation of Roundness-based Estimation

Some may argue that traders on various crypto exchanges are heterogeneous with different algorithmic trading strategies. Therefore, the estimation of wash trade percentage in equation (5) may be distorted in exchanges that have a more significant portion of algorithm trading. If the abnormal unrounded volume is partially inflated by authentic algorithm trading, then our estimates should be viewed more as upper bounds of wash trading.

---

[30] OKEx was highlighted in BTI report as an exchange heavily engaged in wash trading. OKEx has since questioned BTI's methodology and argued that BTI's use of "retail-oriented parameters such as website/mobile traffic" in its research is "an apple-to-orange comparison" (Huillet,2019). In our sample, OKEx indeed fails 20% of all our tests and has an estimated wash trading that is 66% of the volume. But relative to Tier-2 exchanges, it does not deserve a special mention for wash trading. This could be an issue with BTI's methodology as argued but could also be that OKEx has taken actions to either reduce wash trading or avoid being detected.

First, as described earlier, trading algorithms are often believed to be close to exchange-agnostic (Alameda, 2019). Moreover, the controls involving exchange-level observables in Table 7 should also help rule out such a possibility, given that the estimates with controls are comparable to the ones without.

But to drive home the validity of our roundness-ratio approach, we use Benford's law and power law to test if our estimation (Section 4.2) is predominantly capturing wash trading. Because Benford's law and power law are universally applicable to both human and bot trades, they should hold for authentic algorithmic trading. On the other hand, if agents use bots to wash trade, it is likely that these laws do not hold. We, therefore, examine whether the two laws hold for unrounded transactions on both regulated and unregulated exchanges.

We first re-examine whether the first-significant-digit distribution in unrounded trades is consistent with Benford's law, shown in the Online Appendix H. In the sample of unrounded trades, the Chi-squared statistics for conformity with Benford's law is similar to the results in the full sample (see Table 2), indicating that inconsistency with Benford's law may be attributed to the manipulative activities in unrounded trades. All regulated exchanges show a Benford's law distribution in the first significant digit of unrounded trades. Unregulated Tier-1 exchanges exhibit similar patter as the regulated exchanges, while 50% of Tier-2 exchanges violate Benford's law in at least one cryptocurrency pair.

We also find that unrounded trades on regulated exchanges satisfy power law (see Online Appendix I), but unrounded trades on a majority of unregulated exchanges fail the tests, indicating that the unrounded trades cannot be predominantly authentic algorithmic trades.

### 4.4   Alternative Measures and Comparisons with Existing Reports

Given the limitation on data access, quantifying wash trading is a daunting task. We cannot assert that our estimates are the gold standard, especially when one believes that traders and algorithmic strategies are different on different crypto exchanges. As such, we provide two complementary metrics that should help convince the readers that wash trading on unregulated exchanges is rampant and economically significant. We also discuss existing estimations from the industry and why ours are likely to be more robust and superior.

We provide an additional certainty measure to capture the extent of an exchange's wash trading. To this end, we calculate the percentage of failure using results from Online Appendix F, shown in Figure 5.[31] In addition, we compare the trade size distribution of unregulated exchanges to regulated exchanges for robustness (Online Appendix G).[32]

[Insert Figure 5]

In general, unregulated Tier-1 exchanges have lower failure rates (on average 20.6% than unregulated Tier-2 exchanges (on average 61.8%). Some Tier-1 exchanges only show mild patterns of wash trading. Wash trading, once found, can damage exchanges' reputations. It is thus not surprising that some of the unregulated Tier-1 firms might have already been following compliance requirements in jurisdictions outside the United States.

Grouped by cryptocurrency, the percentage of failed tests (wash trading certainty) is the highest in XRP trades (54.2%), followed by BTC (47.4%), LTC (47.0%), and ETH (42.3%).

[Insert Table 8]

We also examine the relationship between failed rates and fractions of wash trades as in Table 8. The percentage of wash trade is positively associated with the percentage of failure at a 1% significance level, a 1% increase in the failure rates corresponds to a 0.597% higher percentage of wash trading. Our estimates for wash trading indeed reflect questionable trading volumes on unregulated exchanges.

We adopt an alternative method to gauge the extent of wash trading using Benford's law. For each exchange, we construct nine counterfactual trade-size distributions based on Benford's law by assuming that all transactions with first-significant-digit X (X being 1 to 9) are authentic, respectively. We then calculate the percentage difference between trade volume estimated by counterfactual first-significant-digit distribution and the volume of actual trade-size distribution. Finally, the extent of wash trade is measured as the median of 9 volume percentage differences to avoid the influence of noise and outliers.

---

[31]Online Appendix F contains three tests concerning universal laws or patterns in traditional financial markets, including the Chi-squared test for Benford's Law, t-test for trade-size clustering, and power-law fitting of the distribution tail. For each exchange, the percentage of failure is measured as the number of failed tests at a 5% significance level over the total number of tests of all four trading assets. Similarly, the percentage of failed tests by cryptocurrency is calculated as the number of failed tests at a 5% significance level over the total number of tests in each of the four cryptocurrency trade pairs we consider.

[32]Online Appendix G conducts the Pearson's Chi-squared test to compare the trade-size distributions of unregulated exchanges to regulated exchanges. We estimate the trade-size percentage in different intervals (e.g. ten thousands, thousands, hundreds, tens, ones, tenths, hundredths, etc.) and its deviation from that in regulated exchanges, whose average are considered as the benchmark. We set the null hypothesis that trade-size distributions are statistically indifferent between unregulated exchanges and the regulated benchmark. Results show that Tier-2 exchanges are more inconsistent with the distribution of regulated exchange than Tier-1 exchanges.

We find that counterfactual distributions of regulated exchanges exhibit little deviation (3.1%) from the actual trade-size distribution, implying the absence of wash trading. However, on average 16.3 % of trade volume is fabricated on unregulated exchanges. Tier-1 unregulated exchanges (12.9%) have a lower fraction of wash trade than Tier-2 unregulated exchanges (18.5%), which is consistent with the previous finding. We report the details in Online Appendix J.

We note that compared with the roundness ratio approach in equation (5), estimates using Benford's law are significantly lower. This does not invalidate the use of roundness ratio as our main estimator because the Benford's based approach would not detect a large fraction of wash trading that contributes to the frequency of all 9 digits being the first significant digits.[33] In that sense, we are essentially underestimating the volume of wash trades. Therefore, the Benford's law approach should be viewed as lower bounds on wash trading, given that heterogeneous traders or strategies across exchanges cannot generate deviations from Benford's law distribution as long as they are authentic.

Although we are the first academic study to quantify wash trading, several industry attempts preceded us. Most notably, Fusaro and Hougan (2019), in their Bitwise report, monitored live trade books from several exchanges' websites and "programmatically read data off the screen" to collect data. They found transactions on unregulated exchanges show larger bid/ask spread, larger order size, and strange volume distribution over time, compared to a few regulated exchanges. While the findings are suggestive, live order books may miss some information due to API trading and iceberg orders, among other issues. Their data are limited and the truncation of the trade-size window is chosen ad hoc. Furthermore, their methods lack formal statistical tests.

Alameda Research, a US-based quant trading firm affiliated with the FTX exchange, addressed the inaccuracy in the Bitwise report in their report in July (Alameda, 2019). They examine the trade history and order book, compare volume correlation with reputable exchanges using self-selected thresholds, assess exchanges liquidity, etc. They assign weighted scores to their detection tests, and then assign 100%, 50%, and 0% wash trade amounts based on the number of tests passed, resulting in imprecise estimates. Their intention was to rank exchanges in terms of wash trading, not to quantify wash trading.

---

[33] In fact, power law and Benford's law only describe the first significant digit or tail distributions instead of the entire distributions of transactions, and are less useful (except for robustness tests) when it comes to quantifying wash trading.

Sylvain Ribes examined the correlation between exchanges' liquidity and the reported volume to evaluate exchanges' volume credibility, although there is no theoretical underpinning for any particular link between slippage and total volume (Ribes, 2018). Blockchain Transparency Institute, a data aggregation website, publishes market surveillance report every quarter since late 2018. They calculate 'clean volume' by conjecturing numbers of visitors which has been criticized by the opacity in their methodology (Huillet, 2019). TokenInsight is not transparent about its methodology used in quantifying wash trading either.

Overall, our analyses not only cover more exchanges and observations than previous studies, but also are transparent and grounded in statistics and behavioral economics. The use of Benford's law, rounding, and power law are well motivated and supported theoretically and empirically. Our tests are systematic and robust to various other factors such as trader heterogeneity across exchanges, as we demonstrated earlier. Given that most of the existing wash trading evidence in the industry is only suggestive and the quantifications imprecise, we contribute by both developing new detection tools that are grounded in universal statistical and behavioral principles and quantifying systematic wash trading in a relatively precise and robust way.

## 5  Wash Trading Incentives, Impacts, and Implications

We now discuss the potential drivers and implications of crypto wash trading. We start with the incentives for wash trading and how it affects the ranking of crypto exchanges. We then analyze the characteristics of exchanges that portend wash trading, explore wash trading's impacts on crypto asset prices, before examining its regulatory and industrial ramifications. Our data limit the extent of the investigation, and we make no causal claims. Nonetheless, the insights gained add to the first canon of knowledge on the topic which is useful for other studies. For example, Amiram, Lyandres, and Rabetti (2021) further examine wash trading in larger panel data and explore how competition interacts with crypto exchanges' operations in both the short and long terms.

Note that wash traders in traditional markets tend to be traders rather than exchanges, yet individuals' wash trades alone cannot fully explain the differences we observe between regulated and unregulated exchanges. Moreover, individuals' cost of wash trading should be related to fees charged and bid-ask spreads (which they have to pay if others cross their orders before they do). But we do not find a systematic correlation between the extent of wash trading and these variables. In contrast, evidence abounds that exchanges themselves wash trade either directly or indirectly. Top executives at crypto exchanges are known to trade on their own exchanges while operating cryptocurrency hedge funds (e.g., Bitfinex'ed, 2017); multiple exchanges have also pleaded guilty of direct wash trading (Sinclair, 2020). Indirect wash trading by the exchanges could be through fee rebates that some exchanges use to incentivize their customers to wash trade. For example, Fcoin rewards platform tokens for trade mining: those individuals who trade more get more rewards in FT tokens.

28Electronic copy available at: https://ssrn.com/abstract=3530220

## 5.1 Wash Trading and Exchange Ranking

Brand awareness and website traffic are two critical factors for customer acquisition, because investors rely on third-party rating or ranking websites to decide which crypto exchange to use. We utilize the proprietary, high-frequency data on exchange ranks and reported trading volumes from CoinMarketCap.com, which most exchanges rely on for referral traffic.[34] To study the incentives for wash trading by crypto exchanges, we first verify the ranking rule of CoinMarketCap using the daily rankings and reported volumes of more than 260 crypto exchanges. Spearman rank-order correlation coefficient is estimated to measure the rank correlation between trade volume and ranking in the CoinMarkCap. The coefficient is -0.995, approaching -1, indicating that ranks and volume are perfectly negatively related (see Figure 6). The rankings of CoinMarketCap are determined by the trade volume of crypto exchanges. Exchanges with larger volumes would rank higher and gain more visibility and visits.

*[Insert Figure 6]*

Exchanges' profit crucially depends on brand awareness and website traffic for customer acquisition, both of which heavily rely on public rankings in broadly recognized data tracking/ranking services or third-party websites such as CoinMarketCap. Our findings support the intuition that to survive the fierce competition, many crypto exchanges naturally wash trade to gain prominence and market share so that the exchange can generate higher profits.[35] Indeed, from Figure 7, we observe that a

---

[34] For instance, according to SimilarWeb reports, one regulated exchange in our sample has around 65% of web traffic referred from CoinMarketCap. On 20 unregulated exchanges, CoinMarketCap is their top 1 referral website and contributes most of web traffics. On 17 unregulated exchanges, web traffic redirection from CoinMarketCap accounts for more than 30% of total web traffic.

[35] Because crypto exchanges are not listed, we do not observe exchanges' revenues and profits. But we can estimate exchanges' profit for the ones that issue their own tokens with utility and dividend functions. Such exchanges periodically use a portion of their operating profit to buyback and destroy tokens from the secondary market (monthly or quarter). We

70% wash trading can move the rank of an exchange up by more than 25 positions relative to its rank in a world without wash trading.

[Insert Figure 7]

## 5.2 Price Impacts of Wash Trading

In Table 9, we examine the effect of wash trading on cryptocurrency prices. Panel A illustrates the relationship between wash trading volumes and weekly returns. Panel B further reports whether wash trading makes the price listed on unregulated exchanges deviate from "fair" prices on regulated exchanges. For each unregulated exchange, price deviation is measured as the log difference between its weekly close price and the average price from regulated exchanges (whose prices are very similar). In both panels, we regress these price indicators on logarithms of estimated wash trade volumes and control for features of exchanges both in contemporaneous and predictive regression specifications. The random-effect model with robust errors is adopted in all regressions based on the Hausman test. We also include currency fixed effect as robustness in both panels.

[Insert Table 9]

As shown in Panel A of Table 9, wash trade volume is positively and significantly associated with the weekly return while lagged wash trade volume has strong negative predictability. The reverse relation with return suggests that higher wash trade volume drives up the contemporaneous price, but the wash-trade effect on price does not last long and price reverses in the following week. What we observe is intuitive: Faking transactions at higher prices can attract more investors who like to chase returns, but arbitrageurs close the pricing gap across exchanges over the next week.

To confirm this intuition, we treat prices on regulated exchanges as "fair" price benchmarks and examine the price deviation of unregulated exchanges against the benchmark. Panel B shows strong and positive relations between wash trade volume and price deviations while controlling for exchange characteristics. In addition, wash trade volume negatively and significantly predicts changes in price deviations in the following week. This is consistent with the notion that speculators

---

manually collect all available buyback reports and token white papers from exchanges' website to compute the value of the tokens bought back or burned. Then with the buyback/profit ratio the exchanges promise (typically described) in the exchange tokens' white papers, we calculate the exchanges' profits. In our sample, UT1, UT3, UT5, UT6, UT8, UT10, U1, U3, U7, U11, and U12 issue exchange tokens and have data available. We find an exchange's profit is positively correlated with both the reported volume and our estimated real volume. In an unreported pooled regression controlling for week fixed effect, the coefficient of log profit on log real volume is 0.85 and significant at a 1% level. We also find that reported CoinMarketCap volume positively and significantly predicts the subsequent week's non-wash-trading volume, consistent with the intuition and empirical findings in Amiram, Lyandres, and Rabetti (2021).

arbitrage away the price differences among various exchanges in the subsequent week and therefore reduce the price deviation.

## 5.3 Determinants of Wash Trading

We first investigate which types of exchanges are more likely to engage in wash trading. We run a cross-sectional regression of the overall fraction of wash trades on an exchange against its characteristics, as shown in Table 10. We include the age of the exchange and all three traffic indicators derived from a series of SimilarWeb reports. Note that number of unique visitors refers to the number of distinct individuals visiting a webpage, which is a close indicator of user number. A smaller number also implies that more visitors may have accessed the exchanges through third-party aggregators or referrals of the ranking websites. The other two indicators are based on each exchange's top 5 traffic geographical origin. We rank all traffic countries in our sample based on GDP and Financial Access.[36] The number of countries ranked at the bottom 15 is counted if these countries appear in the Top 5 traffic countries for crypto exchange.

*[Insert Table 10]*

From Table 10, we observe a negative relationship between the age of exchange and the fraction of wash trades, statistically significant at a 1% level. Moreover, the adjusted $R^2$ is 28.4% in Model 1, implying that the age of exchange is one leading factor correlated with the decision to wash trade. Newly established exchanges are more eager to wash trade since it is a shortcut to increase brand awareness and acquire clients. In addition, the number of unique visitors is negatively associated with wash trading, indicating that exchanges with fewer unique visitors have higher fractions of wash trade.

In fact, unregulated exchanges more than five years old on average wash trade 48.12% of the reported volume as compared with 82.89% for unregulated exchanges no more than five years old; those with more than ten thousand unique users on average wash trade 61.32% of the reported volume as compared with 83.86% for those with no more than ten thousand users. These findings are consistent with the economic incentives of wash trading, and with practitioners' belief that the large exchanges have a reputational consideration to keep things above board and to get it right (Rodgers, 2019).

---

[36] We extract 2016-2018 GDP and financial access data from the World Bank Databank. The measurement of finance access includes the number of commercial bank branches (per 100000 adults), account ownership at a financial institution, and the number of ATM (per 100000 adults). The average value of GDP and financial access measurement is used to rank all traffic countries in our sample.

The insignificant relationship with traffic country indicators implies that the extent of exchanges' wash trading may not vary across countries. We expect exchanges that rely more on referral traffic to have more incentives for wash trading. But this does not show up in our data, either due to the short sampling period or due to the fact that many exchanges may not actively monitor the sources for their web traffic.

Next, we investigate how market dynamics affect wash trading. Table 11 presents a panel regression of wash trade volumes on lagged "true" cryptocurrency weekly return and volatility, which are obtained from the third-party composite price index on CoinMarketCap.[37]

*[Insert Table 11]*

In Table 11, lagged cryptocurrency returns positively predict wash trade volume, while lagged volatility shows a strong negative prediction. In other words, misbehaving crypto exchanges tend to increase wash trading volumes when the market experience recent positive returns or decreases in volatility in the past one or two weeks. Price increases could draw retail investors' attention and encourage speculation. Therefore, crypto exchanges are incentivized to pump up volumes to vie for better ranking and more clients. In addition, decreased volatility reduces the potential costs of wash trading (wash trading risks of capital loss in a volatile market). Therefore, lower volatility can lead to higher wash trading activities.

## 5.4 Regulation's Effects and Implications for Policy and Industry Practice

Concerning regulation, what should we take away from the extensive evidence and large scale of crypto wash trading? Evidently, the supposedly decentralized crypto ecosystems do have centralized players such as the exchanges, which are prone not only to being hacked but also to manipulative behavior. This casts shadows over the industry's development, adding to what the critics have voiced about the limitation of the technology and the fraudulent nature of the industry (Roubini, 2018). Our findings add new insights concerning the role of regulation.

We demonstrate that regulated and unregulated exchanges exhibit vastly divergent trading patterns.[38] Without claiming causality, we offer three potential interpretations of the results. First, as

---

[37] Note that the weekly volatility is calculated using daily returns in the week. All regressions employ random effects with robust errors.
[38] Why do investors trade on unregulated exchanges? Most exchanges started as unregulated and regulation was only introduced gradually. Many investors were unaware of wash trading until 2019, and do not treat regulatory status as their primary decision variable, especially if they have already been trading on an exchange. Customer acquisitions by

we describe in Section 3 and Appendix A, regulated exchanges are directly required to follow the regulation, and violations are severely punished (BitLicense, 2015). The centralized nature of these exchanges, while ironic when we consider the origins of blockchains and decentralized finance, does make direct inspections and the enforcement of regulation on crypto exchanges much more feasible than on other (often anonymous) agents. Second, it is possible that compliance with regulation is costly but does not affect wash trading incentives directly. Some firms simply get a license to signal their quality (e.g., Spence, 1978). This is inconsistent with the observation that after acquiring the license, regulated exchanges still do not wash trade. Third, it is possible that some unobserved exchange characteristics cause the exchange to refrain from wash trading and acquire licenses at the same time. Such a screening function is plausible and would imply that by observing which exchanges are regulated, traders can tell whether wash trading takes place on a particular exchange.

Perhaps contrary to common beliefs, the five regulated spot market exchanges under BitLicense only constitute 0.8% of the total transaction volume in the cryptocurrency market based on CoinMarketCap data. This implies that wash trading on unregulated exchanges is a first-order problem and much more has to be done in terms of regulation. Towards this end, we offer an initial set of tools to convincingly unveil wash trading and to combat non-compliant and unethical behaviors. Regulatory tools and policy have to be adaptive and our statistical tests could become outdated once sophisticated wash traders incorporate them into their strategies. Nevertheless, we believe that the benefits of greater transparency, proper regulation, and close public monitoring that we touch upon are enduring.

## 6   Conclusion

The nascency of the cryptocurrency industry provides a unique setting in which we observe both regulated and unregulated exchanges that are influential. We show that many unregulated crypto exchanges are engaged in excessive wash trading. Specifically, first-digit distributions of trade size follow Benford's law for regulated exchanges, whereas nearly 30% of unregulated exchanges show violations. Furthermore, regulated exchanges show apparent trade clustering at round sizes and a high level of transaction roundness; for unregulated exchanges, the levels of roundness are generally low and the trade-size clustering phenomenon is less prominent. Finally, regulated exchanges display power-law decay with tail exponents in the Pareto–Lévy range, consistent with regularity in

---

unregulated and regulated exchanges are also thus far centered around various promotions, fee cut, reputation within the industry, perceived liquidity, etc.

financial markets; in contrast, 20% of Tier-1 and 75 % of Tier-2 exchanges fail to follow Pareto–Lévy law in trade-size distribution of any cryptocurrency.

We estimate the average wash trading to be 53.4% of trading on unregulated Tier-1 exchanges and 81.8% on Tier-2 exchanges and provide several robustness and validation tests. We further provide suggestive evidence that wash trading inflates exchange rankings and cryptocurrency prices, in addition to being significantly predicted by market signals such as past cryptocurrency prices and volatility and exchange characteristics such as exchange age and userbase. As the first comprehensive study of the pervasive crypto wash trading, our paper not only provides a cautionary tale to regulators around the globe but also reminds the readers of the disciplining or screening effects of regulation in emerging industries, the importance of using wash-trading-adjusted volume in certain empirical studies, and the utility of statistical tools and behavioral benchmarks for forensic finance and fraud detection.

# Table 1. Information on Crypto Exchanges

Table 1 summarizes information on crypto exchanges in the data set. Regulated exchanges are those that are certified and regulated by the New York State Department of Financial Services. Unregulated exchanges are categorized into unregulated Tier-1 and unregulated Tier-2 exchanges based on website traffic ranks. Exchange age is the duration from an exchange's establishment date to July 2019. Exchanges are categorized into three groups based on their length of survival: "more than 5 years," "between 2 and 5 years," and "less than 2 years". Trade volume is calculated as the sum of all transactions involving the four selected cryptocurrency pairs, i.e., BTC, ETH, LTC, and XRP, all against U.S. dollars. SimilarWeb rankings are based on the SimilarWeb report over the period from Aug 2019 to Oct 2019 https://www.similarweb.com/. Alexa's ranking is accessed through https://www.alexa.com/siteinfo in Nov/15/2019. CoinMarketCap ranking is based on daily trade volume, reported on https://www.coinmarketcap.com/ daily averaged during the sample period.

| Exchange Code | Exchange Age | Trade Volume ($mil) | Ranking by Web Traffic | | | Ranking by Trade Volume |
|---|---|---|---|---|---|---|
| | | | SimilarWeb Average Rank in the Investment Section | SimilarWeb Average Number of Monthly Visits (millions) | Alexa Average Rank among all Websites | CoinMarketCap |
| **Panel A Regulated exchanges** | | | | | | |
| R1 | ≥ 5 year | 1466 | 473 | 1.872 | 14297 | 63.7 |
| R2 | ≥ 5 year | 15212 | 17 | 20.678 | 2254 | 50.3 |
| R3 | ≥ 5 year | 1568 | 1418.5 | 0.487 | 23950 | 99.2 |
| **Panel B Unregulated Tier-1 exchanges** | | | | | | |
| UT1 | 2year ≤ A<5year | 41936 | 21 | 18.770 | 1630 | 10.5 |
| UT2 | ≥ 5 year | 434 | 276 | 2.983 | 5960 | 89.9 |
| UT3 | ≥ 5 year | 11175 | 345 | 2.57 | 9683 | 59.5 |
| UT4 | ≥ 5 year | 34157 | 498.5 | 1.363 | 9815 | 27.9 |
| UT5 | ≥ 5 year | 38789 | 285.5 | 1.673 | 8379 | 22.7 |
| UT6 | < 2year | 4005 | 255.5 | 1.879 | 8663 | 55.2 |
| UT7 | ≥ 5 year | 545 | 699 | 0.394 | 13357 | 53.3 |
| UT8 | ≥ 5 year | 24646 | 633 | 1.224 | 3636 | 14.5 |
| UT9 | ≥ 5 year | 975 | 38 | 2.146 | 768 | 95.6 |
| UT10 | ≥ 5 year | 18452 | 517.5 | 1.449 | 5231 | 30.0 |
| **Panel C Unregulated Tier-2 exchanges** | | | | | | |
| U1 | < 2year | 7805 | 17322 | 0.032 | 81142 | 29.9 |
| U2 | < 2year | 30997 | N/A | 0.260 | 3684 | 19.0 |
| U3 | 2year ≤ A<5year | 3464 | 4926.5 | 0.096 | 19860 | 16.1 |
| U4 | < 2year | 50944 | 2594 | 0.234 | 30210 | 10.2 |
| U5 | < 2year | 14534 | 5928.5 | 0.031 | 363745 | 46.6 |
| U6 | 2year ≤ A<5year | 52741 | 6735 | 0.092 | 6422 | 16.0 |
| U7 | < 2year | 34624 | 2770 | 0.265 | 6306 | 11.9 |
| U8 | < 2year | 21848 | 1818.5 | 0.092 | 100223 | 15.0 |
| U9 | 2year ≤ A<5year | 52 | 961.5 | 0.919 | 37634 | 90.0 |
| U10 | < 2year | 2756 | 11567 | 0.007 | 1684659 | 6.6 |
| U11 | < 2year | 32305 | 3403.5 | 0.190 | 1714 | 16.8 |
| U12 | < 2year | 16035 | 3243 | 0.313 | 22780 | 30.8 |
| U13 | < 2year | 2612 | 2316.5 | 0.342 | 28739 | 30.4 |
| U14 | 2year ≤ A<5year | 16668 | 10350.5 | 0.032 | 53000 | 21.3 |
| U15 | < 2year | 23525 | 3061.5 | 0.188 | 1858 | 16.0 |
| U16 | ≥ 5 year | 2013 | 1096.5 | 1.065 | 2808 | 73.7 |

## Table 2. Chi-squared Test for Conformity with Benford's Law

Table 2 presents the Pearson's Chi-squared statistics. The results show whether trade-size distributions of exchanges are consistent with the distribution of Benford's law. Results of four trading pairs are reported, including BTC/USD, ETH/USD, LTC/USD, and XRP/USD. Regulated exchanges are those that are certified and regulated by the New York State Department of Financial Services. Unregulated exchanges are categorized into unregulated Tier-1 and unregulated Tier-2 exchanges based on website traffic ranks. $\chi^2$ statistics and *p*-value are reported in the table. ***, **, and * denote the statistical significance levels at 1%, 5% and 10%, respectively.

| Exchange Code | BTC/USD $\chi^2$ | *p*-value | ETH/USD $\chi^2$ | *p*-value | LTC/USD $\chi^2$ | *p*-value | XRP/USD $\chi^2$ | *p*-value |
|---|---|---|---|---|---|---|---|---|
| **Panel A Regulated exchanges** | | | | | | | | |
| R1 | 1.647 | 0.990 | 1.639 | 0.990 | 4.905 | 0.768 | 11.487 | 0.176 |
| R2 | 2.736 | 0.950 | 2.767 | 0.948 | 3.218 | 0.920 | 2.189 | 0.975 |
| R3 | 3.304 | 0.914 | 0.698 | 1.000 | 1.969 | 0.982 | NA | NA |
| **Panel B Unregulated Tier-1 exchanges** | | | | | | | | |
| UT1 | 2.495 | 0.962 | 4.113 | 0.847 | 4.645 | 0.795 | 7.205 | 0.515 |
| UT2 | 1.464 | 0.993 | 2.620 | 0.956 | 6.117 | 0.634 | 0.748 | 0.999 |
| UT3 | 29.501*** | 0.000 | 5.349 | 0.720 | 7.157 | 0.520 | 47.121*** | 0.000 |
| UT4 | 6.329 | 0.610 | 3.833 | 0.872 | 7.641 | 0.469 | 1.482 | 0.993 |
| UT5 | 6.832 | 0.555 | 3.104 | 0.928 | 1.094 | 0.998 | 0.468 | 1.000 |
| UT6 | 5.969 | 0.651 | 4.100 | 0.848 | 7.386 | 0.496 | 7.790 | 0.454 |
| UT7 | 17.223** | 0.028 | 4.823 | 0.776 | NA | NA | 3.644 | 0.888 |
| UT8 | 2.601 | 0.957 | 1.956 | 0.982 | 3.724 | 0.881 | 4.230 | 0.836 |
| UT9 | 3.228 | 0.919 | 7.886 | 0.445 | 2.454 | 0.964 | 14.219* | 0.076 |
| UT10 | 2.815 | 0.945 | 0.069 | 1.000 | 0.813 | 0.999 | 0.541 | 1.000 |
| **Panel C Unregulated Tier-2 exchanges** | | | | | | | | |
| U1 | 0.548 | 1.000 | 0.949 | 0.999 | NA | NA | NA | NA |
| U2 | 24.261*** | 0.002 | 16.677** | 0.034 | 6.505 | 0.591 | 4.371 | 0.822 |
| U3 | 4.660 | 0.793 | 19.569** | 0.012 | 3.396 | 0.907 | 4.490 | 0.810 |
| U4 | 1.360 | 0.995 | 2.468 | 0.963 | 0.673 | 1.000 | 0.723 | 0.999 |
| U5 | 50.614*** | 0.000 | 8.254 | 0.409 | 124.881*** | 0.000 | 39.69*** | 0.000 |
| U6 | 0.399 | 1.000 | 0.064 | 1.000 | NA | NA | NA | NA |
| U7 | 5.088 | 0.748 | 23.086*** | 0.003 | 60.516*** | 0.000 | 15.300* | 0.054 |
| U8 | 114.788*** | 0.000 | 141.768*** | 0.000 | 31.068*** | 0.000 | 57.021*** | 0.000 |
| U9 | 63.022*** | 0.000 | 122.298*** | 0.000 | NA | NA | 71.949*** | 0.000 |
| U10 | 10.771 | 0.215 | 4.662 | 0.793 | 12.325 | 0.137 | 26.135*** | 0.001 |
| U11 | 2.430 | 0.965 | 7.140 | 0.522 | 4.115 | 0.847 | 7.602 | 0.473 |
| U12 | 0.544 | 1.000 | 0.122 | 1.000 | 1.042 | 0.998 | 14.676* | 0.066 |
| U13 | 1.157 | 0.997 | 2.583 | 0.958 | 11.614 | 0.169 | 4.815 | 0.777 |
| U14 | 0.678 | 1.000 | 23.351*** | 0.003 | 109.944*** | 0.000 | 26.835*** | 0.001 |
| U15 | 2.240 | 0.973 | 0.536 | 1.000 | 0.703 | 1.000 | 2.249 | 0.972 |
| U16 | 1.695 | 0.989 | 0.924 | 0.999 | 1.317 | 0.995 | 0.577 | 1.000 |

## Table 3. Student's *t*-tests for Trade-size Clustering

Table 3 reports the results of *t*-test analysis for the trade size-clustering effect on sampling exchanges. Regulated exchanges are those that are certified and regulated by the New York State Department of Financial Services. Unregulated exchanges are categorized into unregulated Tier-1 and unregulated Tier-2 exchanges based on website traffic ranks. Trading history data of four cryptocurrencies are tested for every exchange separately, including BTC/USD, ETH/USD, LTC/USD, and XRP/USD. The test aims to examine whether trade frequencies at round sizes are higher than the rest of the observation window. Two sets of tests are carried out with different testing points and observation windows: multiples of 100 units with a window radius 50 (100X-50, 100X+50), and multiples of 500 units with a window radius 100 (500X-100, 500X+100). A positive difference indicates that frequency at round size is higher than the rest within the observation window, therefore suggests trade-size clustering. Differences and *t*-statistics are reported in the table. \*\*\*, \*\*, and \* denote positive difference and the statistical significance levels at 1%, 5%, and 10%, respectively.

**Observation range: Multiples of 100 units (100X-50, 100+50)**

| Code | BTC/USD Difference | *t* statistics | ETH/USD Difference | *t* statistics | LTC/USD Difference | *t* statistics | XRP/USD Difference | *t* statistics |
|---|---|---|---|---|---|---|---|---|
| **Panel A Regulated exchanges** | | | | | | | | |
| R1 | 0.091*** | 14.490 | 0.112*** | 12.280 | 0.160*** | 10.767 | 0.063*** | 6.726 |
| R2 | 0.089*** | 14.875 | 0.135*** | 15.647 | 0.109*** | 8.945 | 0.032*** | 2.955 |
| R3 | 0.125*** | 13.655 | 0.119 | 9.713 | 0.203*** | 8.284 | NA | NA |
| **Panel B Unregulated Tier-1 exchanges** | | | | | | | | |
| UT1 | 0.188*** | 16.993 | 0.226*** | 20.740 | 0.179*** | 9.310 | 0.005 | 0.540 |
| UT2 | 0.026* | 1.926 | 0.039** | 2.327 | 0.065*** | 2.943 | 0.076*** | 3.952 |
| UT3 | 0.100*** | 12.654 | 0.078*** | 8.655 | 0.110*** | 6.696 | 0.076*** | 5.681 |
| UT4 | 0.005 | 1.073 | -0.002 | -0.568 | 0.004 | 0.644 | -0.005 | -0.556 |
| UT5 | 0.128*** | 16.895 | 0.083*** | 14.442 | 0.104*** | 8.003 | 0.010 | 1.116 |
| UT6 | -0.015 | -2.668 | -0.001 | -0.081 | -0.003 | -0.089 | -0.014 | -1.379 |
| UT7 | 0.088*** | 6.854 | 0.057*** | 3.685 | NA | NA | 0.132*** | 6.498 |
| UT8 | 0.082*** | 12.620 | 0.067*** | 10.614 | 0.047*** | 5.289 | 0.009 | 0.903 |
| UT9 | 0.084*** | 10.192 | 0.060*** | 5.782 | 0.101*** | 4.018 | 0.054** | 2.570 |
| UT10 | -0.013 | -4.119 | -0.016 | -18.635 | -0.030 | -9.173 | -0.020 | -16.206 |
| **Panel C Unregulated Tier-2 exchanges** | | | | | | | | |
| U1 | -0.016 | -86.208 | -0.022 | -7.374 | NA | NA | NA | NA |
| U2 | -0.015 | -24.733 | -0.014 | -12.297 | -0.017 | -27.701 | -0.017 | -34.675 |
| U3 | 0.030*** | 7.110 | 0.029*** | 3.687 | -0.002 | -0.131 | -0.083 | -2.264 |
| U4 | -0.008 | -5.629 | -0.015 | -5.415 | -0.012 | -2.601 | -0.008 | -1.019 |
| U5 | 0.073*** | 6.573 | -0.027 | -7.279 | -0.015 | -13.844 | -0.014 | -11.199 |
| U6 | -0.020 | -33.174 | -0.022 | -52.875 | NA | NA | NA | NA |
| U7 | 0.019* | 1.952 | 0.096*** | 9.019 | 0.058*** | 9.982 | -0.017 | -15.221 |
| U8 | -0.001 | -0.341 | 0.035*** | 6.552 | -0.005 | -0.804 | -0.008 | -1.207 |
| U9 | 0.106** | 2.313 | 0.032 | 1.038 | NA | NA | -0.022 | -0.450 |
| U10 | -0.004 | -5.622 | -0.015 | -11.549 | -0.016 | -12.730 | -0.015 | -22.775 |
| U11 | 0.259*** | 20.279 | 0.123*** | 31.466 | 0.111*** | 15.258 | -0.017 | -16.156 |
| U12 | -0.015 | -13.164 | -0.014 | -15.846 | -0.021 | -15.304 | -0.035 | -3.158 |
| U13 | 0.034*** | 3.411 | 0.061*** | 8.316 | 0.094*** | 5.662 | 0.083*** | 6.503 |
| U14 | -0.032 | -22.436 | -0.021 | -33.123 | -0.036 | -16.175 | -0.033 | -2.149 |
| U15 | -0.015 | -8.266 | -0.015 | -8.765 | -0.018 | -35.684 | -0.017 | -30.582 |
| U16 | 0.243*** | 20.575 | 0.019** | 2.354 | 0.018* | 1.753 | 0.004 | 0.333 |

**Observation range: Multiples of 500 units (500X-100, 500X +100)**

| Code | BTC/USD Difference | t statistics | ETH/USD Difference | t statistics | LTC/USD Difference | t statistics | XRP/USD Difference | t statistics |
|---|---|---|---|---|---|---|---|---|
| **Panel A Regulated exchanges** | | | | | | | | |
| R1 | 0.203*** | 15.193 | 0.271*** | 15.533 | 0.248*** | 7.904 | 0.166*** | 7.849 |
| R2 | 0.195*** | 16.758 | 0.290*** | 18.503 | 0.206*** | 9.965 | 0.137*** | 5.893 |
| R3 | 0.266*** | 13.145 | 0.310*** | 13.376 | 0.331*** | 7.750 | NA | NA |
| **Panel B Unregulated Tier-1 exchanges** | | | | | | | | |
| UT1 | 0.354*** | 25.223 | 0.391*** | 35.160 | 0.393*** | 16.171 | 0.083*** | 3.529 |
| UT2 | 0.096*** | 3.000 | 0.102*** | 2.898 | 0.114 | 1.691 | 0.137*** | 3.544 |
| UT3 | 0.221*** | 13.626 | 0.193*** | 12.202 | 0.236*** | 7.838 | 0.197*** | 6.004 |
| UT4 | 0.039*** | 2.978 | 0.033*** | 3.572 | 0.039** | 2.086 | 0.035 | 1.602 |
| UT5 | 0.257*** | 24.010 | 0.147*** | 19.769 | 0.198*** | 10.850 | 0.059*** | 3.018 |
| UT6 | -0.018 | -2.342 | 0.024 | 0.889 | 0.069 | 0.960 | -0.030 | -1.427 |
| UT7 | 0.185*** | 5.603 | 0.171*** | 4.938 | NA | NA | 0.247*** | 5.746 |
| UT8 | 0.139*** | 16.418 | 0.105*** | 13.011 | 0.077*** | 5.647 | 0.035** | 2.012 |
| UT9 | 0.163*** | 6.312 | 0.159*** | 7.099 | 0.239*** | 4.518 | 0.096*** | 2.768 |
| UT10 | -0.010 | -2.025 | -0.009 | -6.041 | -0.029 | -3.679 | -0.013 | -7.457 |
| **Panel C Unregulated Tier-2 exchanges** | | | | | | | | |
| U1 | -0.008 | -45.062 | -0.014 | -2.571 | NA | NA | NA | NA |
| U2 | -0.007 | -18.615 | -0.002 | -0.596 | -0.009 | -10.838 | -0.009 | -12.036 |
| U3 | 0.007 | 1.122 | 0.041** | 2.366 | -0.055 | -1.133 | -0.070 | -0.843 |
| U4 | -0.005 | -3.509 | -0.001 | -0.142 | 0.006 | 0.451 | -0.001 | -0.096 |
| U5 | -0.009 | -3.261 | -0.014 | -4.028 | -0.006 | -3.890 | -0.006 | -8.531 |
| U6 | -0.014 | -11.815 | -0.012 | -17.525 | NA | NA | NA | NA |
| U7 | 0.079** | 2.078 | 0.246*** | 15.485 | 0.018* | 2.008 | -0.009 | -7.708 |
| U8 | 0.006 | 1.333 | 0.030*** | 3.498 | 0.000 | -0.022 | 0.003 | 0.415 |
| U9 | 0.182** | 2.880 | 0.070 | 1.154 | NA | NA | 0.059 | 0.602 |
| U10 | -0.002 | -6.491 | -0.007 | -16.342 | NA | NA | NA | NA |
| U11 | 0.369*** | 11.156 | 0.061*** | 9.883 | 0.062*** | 5.522 | -0.008 | -13.686 |
| U12 | -0.001 | -0.743 | -0.008 | -12.134 | -0.012 | -8.184 | NA | NA |
| U13 | 0.150*** | 5.935 | 0.098*** | 6.720 | 0.054*** | 2.845 | 0.155*** | 6.923 |
| U14 | -0.020 | -11.980 | -0.012 | -13.575 | -0.022 | -9.611 | 0.001 | 0.120 |
| U15 | -0.004 | -0.622 | -0.001 | -0.185 | -0.009 | -10.539 | -0.008 | -15.631 |
| U16 | 0.219*** | 8.589 | 0.080*** | 4.489 | 0.051** | 2.499 | 0.036 | 1.442 |

# Table 4. Power-law Fitting

Table 4 presents the results of power-law fitting on sample exchanges. Regulated exchanges are those that are certified and regulated by the New York State Department of Financial Services. Unregulated exchanges are categorized into unregulated Tier-1 and unregulated Tier-2 exchanges based on website traffic ranks. Trading history data of four cryptocurrencies are tested for every exchange separately, including BTC/USD, ETH/USD, LTC/USD, and XRP/USD. Ordinary Least Square (OLS) and Maximum Likelihood Estimation (MLE), are applied for the estimation of scaling parameters $\hat{\alpha}_{OLS}$ and $\hat{\alpha}_{Hill}$, respectively.[39] We also check whether the estimated parameters are within the Pareto–Lévy range (1<α<2) and mark "Y" if both exponents lie within the Pareto–Lévy range.

| Exchange Code | BTC/USD | | | ETH/USD | | | LTC/USD | | | XRP/USD | | |
|---|---|---|---|---|---|---|---|---|---|---|---|---|
| | $\hat{\alpha}_{OLS}$ | $\hat{\alpha}_{Hill}$ | Pareto–Lévy (1<α<2) | $\hat{\alpha}_{OLS}$ | $\hat{\alpha}_{Hill}$ | Pareto–Lévy (1<α<2) | $\hat{\alpha}_{OLS}$ | $\hat{\alpha}_{Hill}$ | Pareto–Lévy (1<α<2) | $\hat{\alpha}_{OLS}$ | $\hat{\alpha}_{Hill}$ | Pareto–Lévy (1<α<2) |
| **Panel A Regulated exchanges** | | | | | | | | | | | | |
| R1 | 1.806 | 1.279 | Y | 1.696 | 1.374 | Y | 1.510 | 1.849 | Y | 1.748 | 1.338 | Y |
| R2 | 1.763 | 1.191 | Y | 1.745 | 1.308 | Y | 1.857 | 1.309 | Y | 1.809 | 1.257 | Y |
| R3 | 1.668 | 1.297 | Y | 1.762 | 1.425 | Y | 1.673 | 1.835 | Y | NA | NA | NA |
| **Panel B Unregulated Tier-1 exchanges** | | | | | | | | | | | | |
| UT1 | 1.669 | 1.209 | Y | 1.795 | 1.436 | Y | 1.836 | 1.411 | Y | 1.960 | 1.430 | Y |
| UT2 | 1.911 | 1.671 | Y | 1.582 | 1.880 | Y | 1.807 | 1.497 | Y | 1.798 | 1.722 | Y |
| UT3 | 1.680 | 1.277 | Y | 1.719 | 1.425 | Y | 1.815 | 1.397 | Y | 1.948 | 1.430 | Y |
| UT4 | 0.620 | 0.663 | N | 0.785 | 0.790 | N | 0.692 | 0.879 | N | 0.552 | 0.803 | N |
| UT5 | 1.750 | 1.089 | Y | 1.842 | 1.505 | Y | 1.871 | 1.447 | Y | 1.966 | 1.651 | Y |
| UT6 | 3.325 | 1.656 | N | 3.014 | 1.609 | N | 4.563 | 5.865 | N | 5.976 | 5.579 | N |
| UT7 | 1.406 | 0.905 | N | 1.494 | 1.358 | Y | NA | NA | NA | 1.282 | 1.231 | Y |
| UT8 | 1.680 | 0.949 | N | 1.675 | 1.020 | Y | 1.863 | 1.320 | Y | 1.812 | 1.212 | Y |
| UT9 | 1.629 | 1.008 | Y | 1.615 | 1.816 | Y | 1.662 | 1.428 | Y | 1.804 | 1.470 | Y |
| UT10 | 1.479 | 1.095 | Y | 1.841 | 1.417 | Y | 1.546 | 0.932 | N | 1.634 | 1.194 | Y |
| **Panel C Unregulated Tier-2 exchanges** | | | | | | | | | | | | |
| U1 | 1.333 | 2.760 | N | 3.345 | 3.941 | N | NA | NA | NA | NA | NA | NA |
| U2 | 5.197 | 7.155 | N | 10.428 | 7.076 | N | 1.739 | 2.046 | N | 2.194 | 1.469 | N |
| U3 | 2.374 | 2.702 | N | 2.035 | 1.546 | N | 2.014 | 4.005 | N | 2.202 | 4.452 | N |
| U4 | 4.546 | 2.724 | N | 4.716 | 3.573 | N | 7.165 | 4.137 | N | 6.356 | 4.157 | N |
| U5 | 2.269 | 1.701 | N | 4.367 | 1.773 | N | 0.641 | 1.299 | N | 8.689 | 4.863 | N |
| U6 | 1.760 | 1.638 | Y | 1.998 | 1.622 | Y | NA | NA | NA | NA | NA | NA |
| U7 | 7.660 | 7.063 | N | 3.598 | 11.444 | N | 14.815 | 11.706 | N | 12.439 | 6.862 | N |
| U8 | 1.020 | 0.952 | N | 1.157 | 0.874 | N | 1.241 | 0.765 | N | 0.656 | 0.650 | N |
| U9 | 1.370 | 3.770 | N | 1.520 | 3.087 | N | NA | NA | NA | 1.486 | 6.373 | N |
| U10 | 4.292 | 7.578 | N | 7.384 | 7.966 | N | 5.049 | 8.802 | N | 10.697 | 13.863 | N |
| U11 | 5.829 | 6.384 | N | 3.639 | 5.961 | N | 3.676 | 4.877 | N | 7.116 | 5.027 | N |
| U12 | 2.854 | 1.728 | N | 1.926 | 1.880 | Y | 1.572 | 1.226 | Y | 1.831 | 2.691 | N |
| U13 | 1.509 | 1.022 | Y | 1.669 | 1.191 | Y | 1.479 | 1.193 | Y | 1.434 | 1.180 | Y |
| U14 | 0.718 | 1.261 | N | 2.031 | 1.237 | N | 1.077 | 1.056 | Y | 6.551 | 10.524 | N |
| U15 | 1.537 | 1.038 | Y | 1.618 | 1.117 | Y | 1.679 | 1.129 | Y | 1.548 | 1.001 | Y |
| U16 | 2.048 | 1.631 | N | 1.925 | 1.954 | Y | 2.173 | 2.430 | N | 2.175 | 2.074 | N |

---

[39] We apply the probability density function to estimate the scaling exponents 1+ α.

**Table 5. Multiple Hypothesis Testing**

Table 5 presents the multiple hypothesis analysis using Fisher's combined probability test for regulated and unregulated exchanges. For each crypto exchange-cryptocurrency pair, p-values of three sets of tests are used to compute combined statistic $\chi^2$, including the Chi-squared test for Benford's Law, t-test for trade-size clustering, and linear fit for power law. In the global hypothesis testing, the null hypothesis, $H_0$, is that trade patterns of crypto exchanges are consistent with universal laws or patterns in traditional financial markets. The null hypothesis is rejected if $\chi^2$ is larger than the critical value 12.592. In the table below, 1 denotes the null hypothesis rejected and 0 otherwise. Panel A, Panel B, and Panel C show summative results in regulated exchanges, Tier-1 unregulated and Tier-2 unregulated exchanges, respectively. Regulated exchanges are those that are certified and regulated by the New York State Department of Financial Services. Unregulated exchanges are categorized into unregulated Tier-1 and unregulated Tier-2 exchanges based on website traffic ranks. For each test, we report four cryptocurrency pairs, BTC, ETH, LTC, and XRP.

| Exchange Code | BTC Combined $\chi^2$ | Reject $H_0$ | ETH Combined $\chi^2$ | Reject $H_0$ | LTC Combined $\chi^2$ | Reject $H_0$ | XRP Combined $\chi^2$ | Reject $H_0$ |
|---|---|---|---|---|---|---|---|---|
| **Panel A Regulated exchanges** | | | | | | | | |
| R1 | 0.009 | 0 | 0.009 | 0 | 0.229 | 0 | 1.509 | 0 |
| R2 | 0.045 | 0 | 0.046 | 0 | 0.072 | 0 | 0.023 | 0 |
| R3 | 0.078 | 0 | 0.000 | 0 | 0.016 | 0 | NA | NA |
| **Panel B Unregulated Tier-1 exchanges** | | | | | | | | |
| UT1 | 0.034 | 0 | 0.144 | 0 | 0.199 | 0 | 0.880 | 0 |
| UT2 | 0.031 | 0 | 0.048 | 0 | 0.398 | 0 | 0.001 | 0 |
| UT3 | 16.000 | 1 | 0.285 | 0 | 0.568 | 0 | 16.000 | 1 |
| UT4 | 16.562 | 1 | 17.209 | 1 | 16.919 | 1 | 17.083 | 1 |
| UT5 | 0.511 | 0 | 0.065 | 0 | 0.002 | 0 | 0.124 | 0 |
| UT6 | 21.047 | 1 | 16.803 | 1 | 17.274 | 1 | 18.803 | 1 |
| UT7 | 3.106 | 0 | 0.220 | 0 | NA | NA | 0.103 | 0 |
| UT8 | 0.038 | 0 | 0.016 | 0 | 0.110 | 0 | 0.332 | 0 |
| UT9 | 0.073 | 0 | 0.703 | 0 | 0.032 | 0 | 2.244 | 0 |
| UT10 | 16.651 | 1 | 16.602 | 1 | 16.603 | 1 | 16.602 | 1 |
| **Panel C Unregulated Tier-2 exchanges** | | | | | | | | |
| U1 | 16.602 | 1 | 32.603 | 1 | NA | NA | NA | NA |
| U2 | 38.000 | 1 | 35.539 | 1 | 17.059 | 1 | 32.772 | 1 |
| U3 | 16.201 | 1 | 19.842 | 1 | 3.499 | 0 | 19.482 | 1 |
| U4 | 32.606 | 1 | 32.635 | 1 | 20.420 | 1 | 17.610 | 1 |
| U5 | 32.000 | 1 | 33.379 | 1 | 48.602 | 1 | 48.602 | 1 |
| U6 | 16.602 | 1 | 16.677 | 1 | NA | NA | NA | NA |
| U7 | 16.275 | 1 | 21.046 | 1 | 32.000 | 1 | 35.137 | 1 |
| U8 | 16.871 | 1 | 16.000 | 1 | 17.349 | 1 | 33.881 | 1 |
| U9 | 16.014 | 1 | 16.145 | 1 | NA | NA | 16.961 | 1 |
| U10 | 33.937 | 1 | 32.804 | 1 | 25.646 | 1 | 25.710 | 1 |
| U11 | 16.031 | 1 | 16.565 | 1 | 16.144 | 1 | 33.252 | 1 |
| U12 | 32.602 | 1 | 16.602 | 1 | 16.604 | 1 | 5.894 | 0 |
| U13 | 0.003 | 0 | 0.037 | 0 | 1.544 | 0 | 0.219 | 0 |
| U14 | 32.602 | 1 | 30.933 | 1 | 32.602 | 1 | 25.126 | 1 |
| U15 | 16.626 | 1 | 16.602 | 1 | 16.602 | 1 | 16.627 | 1 |
| U16 | 16.010 | 1 | 0.009 | 0 | 16.041 | 1 | 16.401 | 1 |

## Table 6. Chi-squared Test for Trade-size Roundness of Unregulated Exchanges

Table 6 presents the results of Pearson's Chi-squared test on the roundness of unregulated exchanges with respect to the regulated exchanges as a benchmark. Regulated exchanges are those that are certified and regulated by the New York State Department of Financial Services. Unregulated exchanges are categorized into unregulated Tier-1 and unregulated Tier-2 exchanges based on website traffic ranks, shown in Panel A and Panel B, respectively. Trading history data of four cryptocurrencies are tested for every exchange separately, including BTC/USD, ETH/USD, LTC/USD, and XRP/USD. The level of roundness is a parameter describing the decimal or integer places of the last non-zero digit. Test results, $\chi^2$ statistics and *p*-values, reveal the difference of distributions between regulated and unregulated exchanges. ***, **, and * denote the statistical significance levels at 1%, 5%, and 10%, respectively.

| Exchange Code | BTC/USD $\chi^2$ | *p*-value | ETH/USD $\chi^2$ | *p*-value | LTC/USD $\chi^2$ | *p*-value | XRP/USD $\chi^2$ | *p*-value |
|---|---|---|---|---|---|---|---|---|
| **Panel A Unregulated Tier-1 exchanges** | | | | | | | | |
| UT1 | 9.545 | 0.145 | 15.013** | 0.020 | 12.18** | 0.032 | 11.993*** | 0.007 |
| UT2 | 3.100 | 0.796 | 11.455* | 0.075 | 9.222 | 0.101 | 13.387*** | 0.004 |
| UT3 | 92.104*** | 0.000 | 8.086 | 0.232 | 5.616 | 0.345 | 51.094*** | 0.000 |
| UT4 | 17.224*** | 0.008 | 13.387** | 0.037 | 7.547 | 0.183 | 11.393*** | 0.010 |
| UT5 | 115.48*** | 0.000 | 11.01* | 0.088 | 14.311** | 0.014 | 9.5** | 0.023 |
| UT6 | 7.909 | 0.245 | 17.469*** | 0.008 | 24.886*** | 0.000 | 16.603*** | 0.001 |
| UT7 | 182.435*** | 0.000 | 16.518** | 0.011 | NA | NA | 49.766*** | 0.000 |
| UT8 | 4.384 | 0.625 | 15.649** | 0.016 | 19.46*** | 0.002 | 12.18*** | 0.007 |
| UT9 | 3.247 | 0.777 | 5.427 | 0.490 | 11.906** | 0.036 | 14.268*** | 0.003 |
| UT10 | 1461.8*** | 0.000 | 692.292*** | 0.000 | 21.797*** | 0.001 | 18.032*** | 0.000 |
| **Panel B Unregulated Tier-2 exchanges** | | | | | | | | |
| U1 | 18.774*** | 0.005 | 32.402*** | 0.000 | NA | NA | NA | NA |
| U2 | 60.923*** | 0.000 | 62.726*** | 0.000 | 28.101*** | 0.000 | 19.651*** | 0.000 |
| U3 | 828.828*** | 0.000 | 85.86*** | 0.000 | 22.242*** | 0.000 | 19.593*** | 0.000 |
| U4 | 1670.819*** | 0.000 | 31.158*** | 0.000 | 32.097*** | 0.000 | 19.747*** | 0.000 |
| U5 | 1668.236*** | 0.000 | 20.761*** | 0.002 | 27.753*** | 0.000 | 19.109*** | 0.000 |
| U6 | 1639.493*** | 0.000 | 24.944*** | 0.000 | NA | NA | NA | NA |
| U7 | 9.569 | 0.144 | 15.481** | 0.017 | 18.705*** | 0.002 | 19.688*** | 0.000 |
| U8 | 740.835*** | 0.000 | 157.443*** | 0.000 | 86.741*** | 0.000 | 18.59*** | 0.000 |
| U9 | 15.455** | 0.017 | 26.838*** | 0.000 | NA | NA | 19.182*** | 0.000 |
| U10 | 1719.65*** | 0.000 | 23.694*** | 0.001 | 32.242*** | 0.000 | 19.796*** | 0.000 |
| U11 | 439.322*** | 0.000 | 101.26*** | 0.000 | 14.106** | 0.015 | 19.458*** | 0.000 |
| U12 | 18.605*** | 0.005 | 28.754*** | 0.000 | 22.785*** | 0.000 | 19.768*** | 0.000 |
| U13 | 26.08*** | 0.000 | 130.687*** | 0.000 | 41.623*** | 0.000 | 34.596*** | 0.000 |
| U14 | 1310.242*** | 0.000 | 34.176*** | 0.000 | 30.144*** | 0.000 | 19.728*** | 0.000 |
| U15 | 1546.727*** | 0.000 | 23.247*** | 0.001 | 29.609*** | 0.000 | 19.592*** | 0.000 |
| U16 | 535.379*** | 0.000 | 55.367*** | 0.000 | 13.247** | 0.021 | 15.288*** | 0.002 |

## Table 7. Estimating the Fraction of Wash Trades

Table 7 reports the pooled regression results of the fraction of wash trading for unregulated exchanges. The regression equation below specifies the relationship between round and unrounded trade volumes.

$$\ln(V_{Unrounded_{it}}) = \alpha + \beta * \ln(V_{Round_{it}}) + \gamma * X_{it} + \epsilon_{it}$$

where $\ln(V_{Round_{it}})$ and $\ln(V_{Unrounded_{it}})$ are the logarithms of round trade volume and unrounded trade volume, respectively, for exchange *i* at week *t*. $X_{it}$ is a vector of exchange characteristics and $\epsilon_{it}$ is an error term. We categorize trading volume into round and unrounded ones by checking if the mantissa of a particular transaction volume is less than 100 base units or not. Exchange characteristics such as age, rank, CoinMarketCap web traffic percentage, and unique visitors are used as control variables. Exchange U2 and U7 do not have data of control variables. The regression coefficients are used as a benchmark to calculate the expected unrounded trading volume, then the fraction of wash trading for each unregulated exchange. Fractions of wash trading are estimated for each cryptocurrency of each exchange (Panel B and C for unregulated Tier 1 and 2 exchanges, respectively) and then aggregated amount (Panel A) using equal- and volume-weighted averages. A thousand bootstrapped samples are used to calculate the standard deviation of wash trading estimates, which we report in brackets.

**Panel A: Aggregated Wash Trading Percentage**

|  | Wash Trade Percentage Without Control Variables | | Wash Trade Percentage With Control Variables | |
| --- | --- | --- | --- | --- |
|  | Equal-weighted Average | Volume-weighted Average | Equal-weighted Average | Volume-weighted Average |
| Unregulated | 70.85 | 77.50 | 60.96 | 71.43 |
| Unregulated Tier-1 | 53.41 | 61.86 | 46.95 | 63.62 |
| Unregulated Tier-2 | 81.76 | 86.26 | 70.96 | 76.96 |

**Panel B: Wash Trading Percentage for Unregulated Tier-1 Exchanges**

| Exchange Code | Wash Trade Percentage No Control | Wash Trade Percentage With Control |
| --- | --- | --- |
| UT1 | 51.76 (1.28) | 46.47 (1.34) |
| UT2 | 51.73 (1.65) | 18.91 (2.34) |
| UT3 | 1.87 (0.52) | 31.34 (2.06) |
| UT4 | 92.60 (0.66) | 89.81 (1.93) |
| UT5 | 44.87 (2.08) | 57.77 (1.69) |
| UT6 | 74.36 (1.30) | 52.96 (6.67) |
| UT7 | 19.02 (1.55) | 3.02 (1.41) |
| UT8 | 66.12 (1.52) | 72.75 (2.02) |
| UT9 | 37.49 (2.46) | 14.94 (2.19) |
| UT10 | 94.31 (0.54) | 81.49 (4.20) |

**Panel C: Wash Trading Percentage for Unregulated Tier-2 Exchanges**

| Exchange Code | Wash Trade Percentage No Control | Wash Trade Percentage With Control |
|---|---|---|
| U1 | 99.99 (0.00) | 99.93(0.01) |
| U2 | 99.36 (0.13) | NA |
| U3 | 72.72 (2.41) | 72.62(2.18) |
| U4 | 95.50 (0.52) | 91.64(1.51) |
| U5 | 89.71 (0.39) | 72.48(2.55) |
| U6 | 98.13 (0.21) | 98.65(0.11) |
| U7 | 82.00 (3.68) | NA |
| U8 | 77.09 (2.17) | 48.62(5.32) |
| U9 | 81.12 (4.21) | 64.99(3.85) |
| U10 | 98.45 (0.09) | 86.12(2.27) |
| U11 | 34.32 (6.57) | 33.63(5.75) |
| U12 | 98.10 (1.07) | 94.79(2.04) |
| U13 | 65.42 (2.12) | 61.71(2.21) |
| U14 | 96.80 (1.10) | 81.24(3.18) |
| U15 | 94.36 (0.48) | 68.66(5.38) |
| U16 | 25.04 (4.49) | 18.42(4.47) |

**Table 8. Failure Rates of the Statistical Tests and the Fraction of Wash Trades**

Table 8 presents the regression analysis of the fraction of the estimated wash trade on the failure percentage of statistical tests. The percentage of failed tests is calculated as the number of failed tests over the total number of tests across cryptocurrencies, including the Chi-squared test for Benford's Law, *t*-test for trade-size clustering, and tail exponents for the power law (Refer to Online Appendix F). *t*-statistics are reported in the brackets. ***, **, and * denote the statistical significance levels at 1%, 5%, and 10%, respectively.

| | Fraction of wash trades in unregulated exchanges |
|---|---|
| Percentage of Failed Tests | 0.597*** |
| | (4.99) |
| Constant | 0.412*** |
| | (4.54) |
| Observations | 26 |
| Adjusted $R^2$ | 35.2% |

## Table 9. Price Impacts of Wash Trading

Table 9 presents the regression analysis on the price impacts of wash trading. In Panel A, the dependent variable is the weekly returns for every cryptocurrency on every exchange. In Panel B, the price deviation is calculated as the (log) difference between the close price of each unregulated exchange and averaged close prices of regulated exchanges at the same time. In both panels, Exchange Age$_t$ is the time span from its establishment to week t for an exchange. CoinMarketCapRank$_t$ is the rank directly obtained from CoinMarketCap. Tier-1 Exchange is a dummy variable which equals 1 if the exchange is unregulated Tier-1 exchange, 0 otherwise. The number of unique visitors refers to the number of distinct visitors recorded during the sample period, derived from SimilarWeb August to October 2019 reports. All models are estimated with random effects based on the Hausman test. Currency fixed effects are included in Model 2, 4, 6, 8, 10 and 12 of Panel A, and Model 2 and 4 of Panel B. *t*-statistics are reported in the brackets. ***, **, and * denote the statistical significance levels at 1%, 5%, and 10%, respectively.

### Panel A: Returns and Wash Trading

| | Weekly return$_t$ | | | | | | | | | | | |
|---|---|---|---|---|---|---|---|---|---|---|---|---|
| | (1) | (2) | (3) | (4) | (5) | (6) | (7) | (8) | (9) | (10) | (11) | (12) |
| (log) wash trade volume$_t$ | 0.001*** | 0.001 | 0.003*** | 0.002*** | | | | | 0.024*** | 0.023*** | 0.024*** | 0.023*** |
| | (2.61) | (1.33) | (3.24) | (3.12) | | | | | (4.75) | (4.68) | (4.66) | (4.61) |
| (log) wash trade volume$_{t-1}$ | | | | | -0.001*** | -0.002*** | -0.002*** | -0.003*** | -0.024*** | -0.024*** | -0.024*** | -0.024*** |
| | | | | | (-2.95) | (-4.80) | (-3.33) | (-4.42) | (-4.83) | (-4.83) | (-4.69) | (-4.70) |
| Exchange Age$_t$ | | | 0.000 | 0.000* | | | 0.000 | 0.000 | | | 0.000 | 0.000 |
| | | | (1.18) | (1.80) | | | (0.34) | (0.78) | | | (0.65) | (1.60) |
| CMC rank$_t$ | | | -0.002 | -0.004 | | | 0.004 | 0.003 | | | 0.000 | -0.002 |
| | | | (-0.33) | (-0.91) | | | (1.10) | (1.12) | | | (0.01) | (-0.87) |
| Tier-1 Exchange | | | -0.000 | 0.000 | | | -0.001 | -0.001 | | | -0.000 | 0.000 |
| | | | (-0.28) | (0.22) | | | (-1.14) | (-1.26) | | | (-0.25) | (0.68) |
| (log) Number of Unique Visitors | | | 0.000*** | 0.000*** | | | -0.000 | -0.000** | | | 0.000 | 0.000 |
| | | | (2.96) | (3.43) | | | (-1.33) | (-2.19) | | | (1.12) | (1.19) |
| Constant | -0.049*** | -0.027*** | -0.083*** | -0.073*** | 0.010 | 0.030*** | 0.036** | 0.065*** | -0.008 | 0.010 | -0.017 | -0.004 |
| | (-5.15) | (-3.11) | (-3.40) | (-3.69) | (1.28) | (3.49) | (2.16) | (3.42) | (-1.14) | (1.46) | (-0.99) | (-0.31) |
| | | | | | | | | | | | | |
| Currency Fixed Effects | N | Y | N | Y | N | Y | N | Y | N | Y | N | Y |
| Observation | 1416 | 1416 | 1328 | 1328 | 1326 | 1326 | 1246 | 1246 | 1305 | 1305 | 1225 | 1225 |
| Overall R$^2$ | 0.1% | 1.0% | 0.4% | 1.3% | 0.1% | 1.1% | 0.2% | 1.2% | 3.1% | 4.0% | 3.3% | 4.1% |

**Panel B: Price Deviations and Wash Trading**

|  | PriceDeviation$_t$ | | PriceDeviation$_{t+1}$ - PriceDeviation$_t$ | |
|---|---|---|---|---|
|  | (1) | (2) | (3) | (4) |
| (log) wash trade volume$_t$ | 0.047*** | 0.041*** | -0.049*** | -0.052*** |
|  | (3.46) | (2.75) | (-4.18) | (-3.51) |
| Exchange Age$_t$ | 0.000 | 0.000* | -0.000 | -0.000 |
|  | (1.52) | (1.81) | (-0.41) | (-0.46) |
| CMC rank$_t$ | 0.005*** | 0.013 | -0.003*** | -0.092 |
|  | (4.26) | (0.16) | (-3.28) | (-0.82) |
| Tier-1 Exchange | 0.029 | -0.018 | -0.097 | 0.020 |
|  | (0.33) | (-1.04) | (-0.89) | (0.97) |
| (log) Number of Unique Visitors | -0.021 | 0.005*** | 0.021 | -0.003*** |
|  | (-1.08) | (3.92) | (1.04) | (-2.99) |
| Constant | -1.172*** | -1.032** | 1.137*** | 1.208*** |
|  | (-3.14) | (-2.53) | (-3.15) | (2.65) |
|  |  |  |  |  |
| Currency fixed effect | N | Y | N | Y |
| Observation | 1328 | 1328 | 1246 | 1246 |
| Overall R2 | 0.7% | 1.1% | 0.4% | 0.4% |

## Table 10. Wash Trading and Exchange Characteristics

Table 10 reports the cross-sectional regression analysis for the relationship between the fraction of overall wash trading volume for an exchange and its characteristics. Exchange age is the span between the establishment date and July 2019, the start of our sample period. The remaining indicators are derived from SimilarWeb August to October 2019 reports. The number of unique visitors refers to the number of distinct visitors recorded during the sampling period. Top 5 traffics from lower GDP countries refers to the number of traffic countries ranked at the bottom 15 countries based on GDP. Top 5 traffics from worst finance access countries denotes the number of traffic countries ranked at the bottom 15 countries based on financial access. GDP and financial access data are obtained from the World Bank DataBank. The rank of countries is based on the average value of GDP and financial access over three years from 2016 to 2018. $t$-statistics are reported in the brackets. ***, **, and * denote the statistical significance levels at 1%, 5%, and 10%, respectively.

| Fraction of wash trades | Unregulated exchange | | |
|---|---|---|---|
| | Model 1 | Model 2 | Model 3 |
| Exchange Age | -0.659*** | | -0.678*** |
| | (-2.99) | | (-3.09) |
| Number of Unique Visitors | | -0.099** | -0.091*** |
| | | (-2.12) | (-3.70) |
| Top 5 Traffics from Lower GDP Countries | | | 3.158 |
| | | | (0.65) |
| Top 5 Traffics from Worst Financial Access Countries | | | 4.984 |
| | | | (0.92) |
| Constant | 94.420*** | 72.995*** | 87.160*** |
| | (11.55) | (11.69) | (8.12) |
| | | | |
| Observations | 26 | 26 | 26 |
| Adjusted $R^2$ | 28.4% | 1.0% | 30.1% |

## Table 11. Influence of Returns and Volatility on Wash Trading Volumes

Table 11 presents the panel regression results for the impact of weekly cryptocurrency returns and volatility on wash trading volumes of unregulated exchanges. The weekly returns and volatility are calculated based on the third-party composite price indexes from CoinMarketCap (CMC). CMC Volatility$_{t-1}$ is the standard deviation of daily returns during week $t$-$1$. Random-effect models with robust errors are used in all regressions. $t$-statistics are reported in the brackets. ***, **, and * denote the statistical significance levels at 1%, 5%, and 10%, respectively.

| (log) Wash Trade Volume$_t$ | (1) | (2) | (3) | (4) | (5) | (6) | (7) |
|---|---|---|---|---|---|---|---|
| Weekly CMC Return$_{t-1}$ | 1.258*** | | 1.444*** | | | | 1.415*** |
| | (7.14) | | (7.68) | | | | (7.16) |
| Weekly CMC Return$_{t-2}$ | | 0.318** | 0.627*** | | | | 0.350** |
| | | (2.09) | (3.95) | | | | (2.22) |
| CMC Volatility$_{t-1}$ | | | | -5.717*** | | -5.636*** | -4.116*** |
| | | | | (-6.06) | | (-6.03) | (-4.35) |
| CMC Volatility$_{t-2}$ | | | | | -2.297** | -2.070** | -3.547*** |
| | | | | | (-2.18) | (-2.00) | (-3.15) |
| (log) Wash Trade Volume$_{t-1}$ | 0.887*** | 0.882*** | 0.886*** | 0.885*** | 0.882*** | 0.884*** | 0.885*** |
| | (48.67) | (47.61) | (47.93) | (50.07) | (47.86) | (49.38) | (48.56) |
| Constant | 2.304*** | 2.386*** | 2.352*** | 2.543*** | 2.459*** | 2.632*** | 2.619*** |
| | (6.62) | (6.71) | (6.64) | (7.21) | (6.80) | (7.19) | (7.10) |
| | | | | | | | |
| Observation | 1305 | 1305 | 1305 | 1305 | 1305 | 1305 | 1305 |
| Overall $R^2$ | 92.9% | 92.7% | 93.0% | 92.9% | 92.8% | 93.0% | 93.2% |

## Figure 1. First-significant-digit Distribution and Benford's Law

Figure 1 displays the first-significant-digit distributions and comparison with Benford's law. R2, UT6, U8, U9, and U14 are five exchanges selected from regulated exchanges, Tier-1 unregulated and Tier-2 unregulated exchanges, respectively. Regulated exchanges are those that are certified and regulated by the New York State Department of Financial Services. Unregulated exchanges are categorized into unregulated Tier-1 and unregulated Tier-2 exchanges based on website traffic ranks. Distributions of four trading pairs are reported in bar charts, including BTC/USD, ETH/USD, LTC/USD, and XRP/USD. Black dots represent distributions derived from Benford's law.

**Panel R: Regulated Exchanges**
**R2**

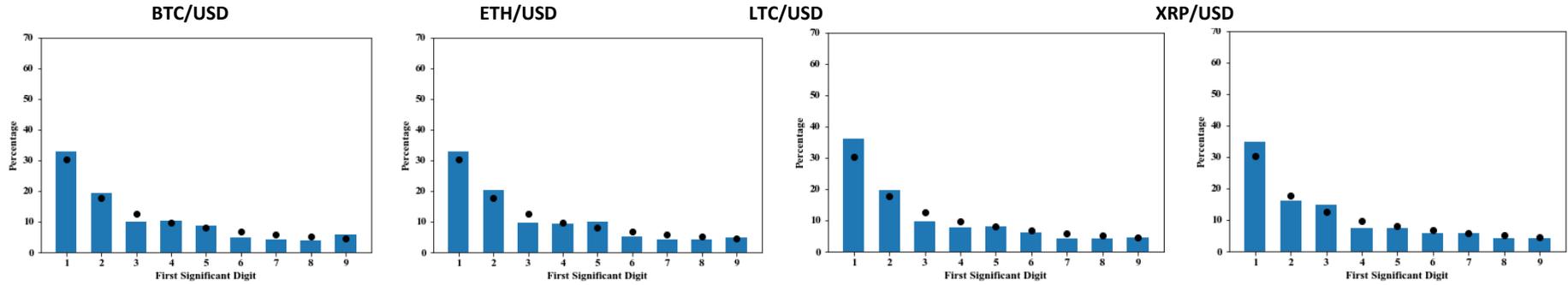

**Panel UT: Unregulated Tier-1 Exchanges**
**UT6**

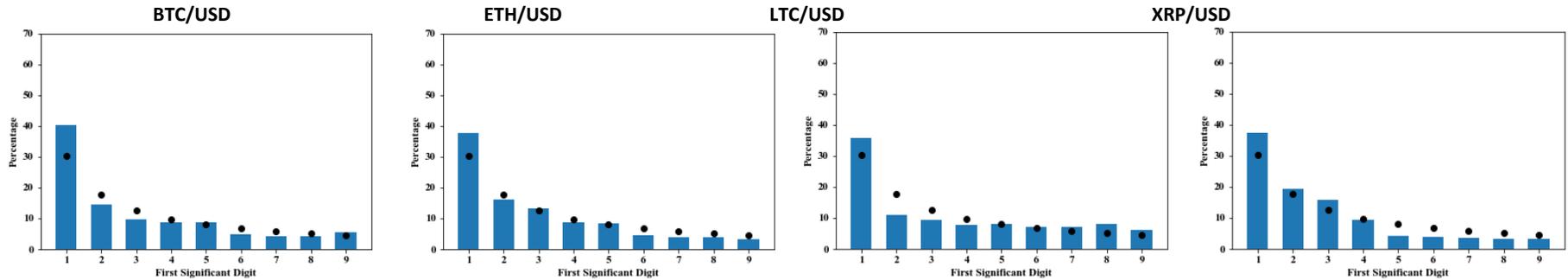

**Panel U: Unregulated Tier-2 Exchanges**

**U8**

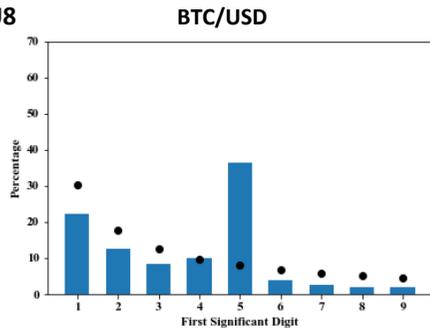
BTC/USD

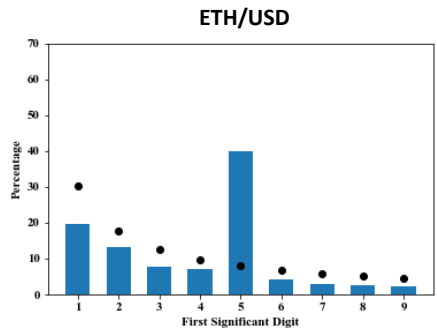
ETH/USD

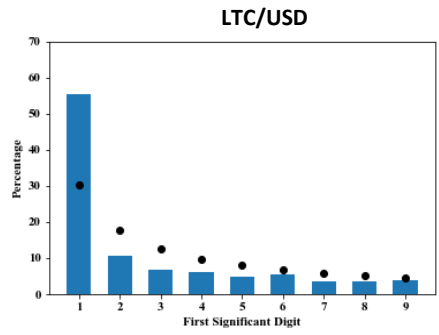
LTC/USD

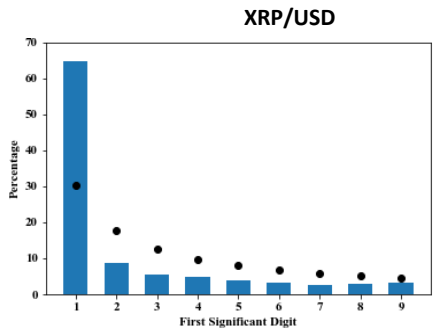
XRP/USD

**U9**

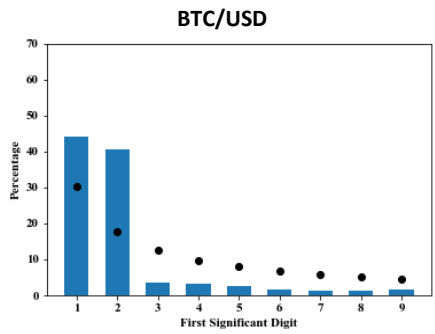
BTC/USD

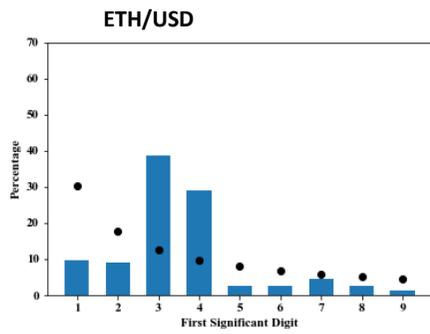
ETH/USD

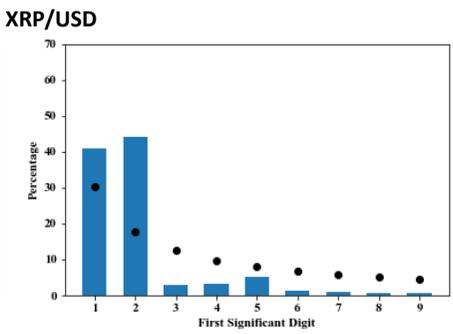
XRP/USD

**U14**

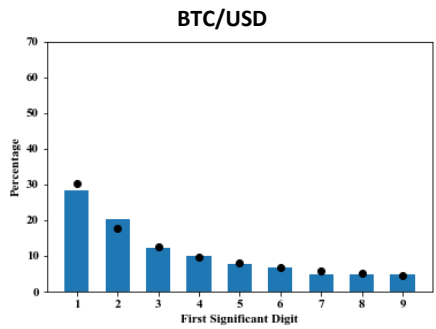
BTC/USD

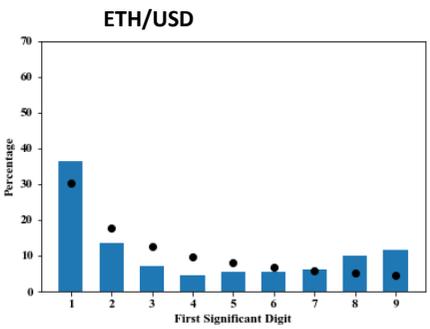
ETH/USD

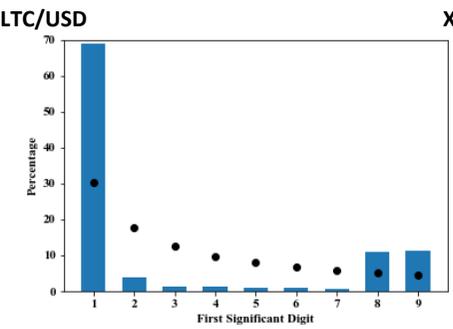
LTC/USD

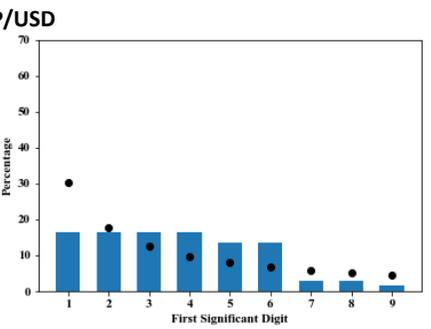
XRP/USD

# Figure 2. Trade-size Clustering

Figure 2 depicts the clustering effect in trade-size distributions histograms on exchanges R2, UT6, U8, U9, and U14. Panel R, Panel UT, and Panel U refer to regulated exchanges, Tier-1 unregulated, and Tier-2 unregulated exchanges, respectively. Regulated exchanges are those that are certified and regulated by the New York State Department of Financial Services. Unregulated exchanges are categorized into unregulated Tier-1 and unregulated Tier-2 exchanges based on website traffic ranks. Four trading pairs, including BTC/USD, ETH/USD, LTC/USD, and XRP/USD, are reported for each exchange separately. Two sets of observation ranges are applied for each trading pair: 0-1 BTC, 0-10 BTC, 0-10 ETH, 0-100 ETH, 0-100 LTC, 0-1000 LTC, 0-10000 XRP, and 0-100000 XPR. In each histogram, we highlight every 5th and 10th bin to illustrate the clustering effect around round trade sizes.

**Panel R: Regulated Exchanges**
R2

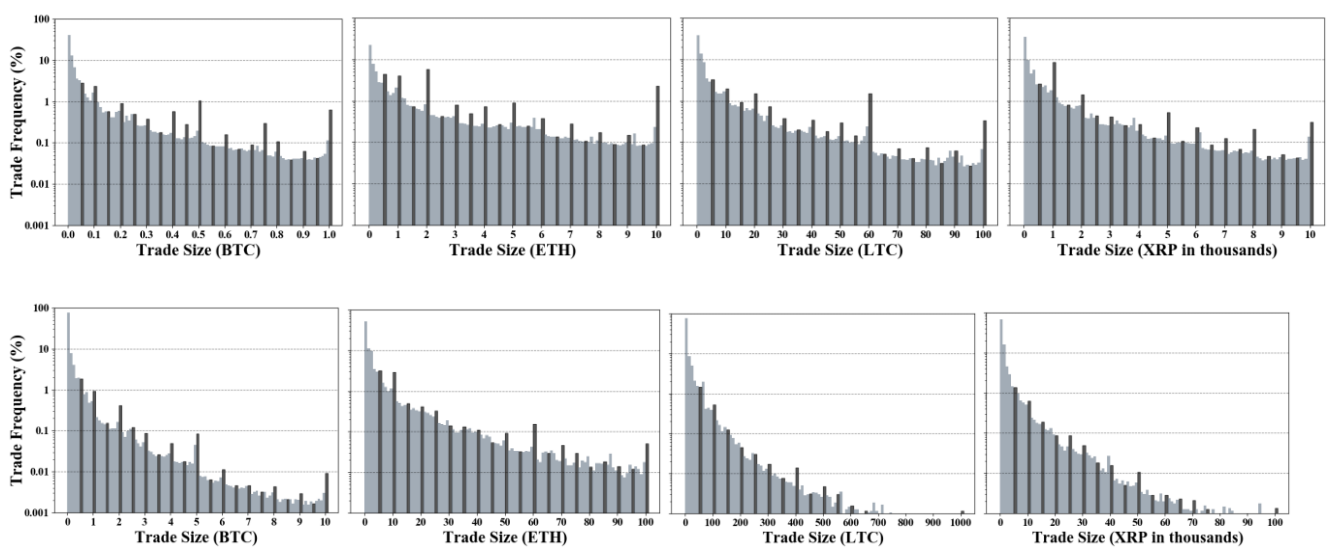

**Panel UT: Unregulated Tier-1 Exchanges**
UT6

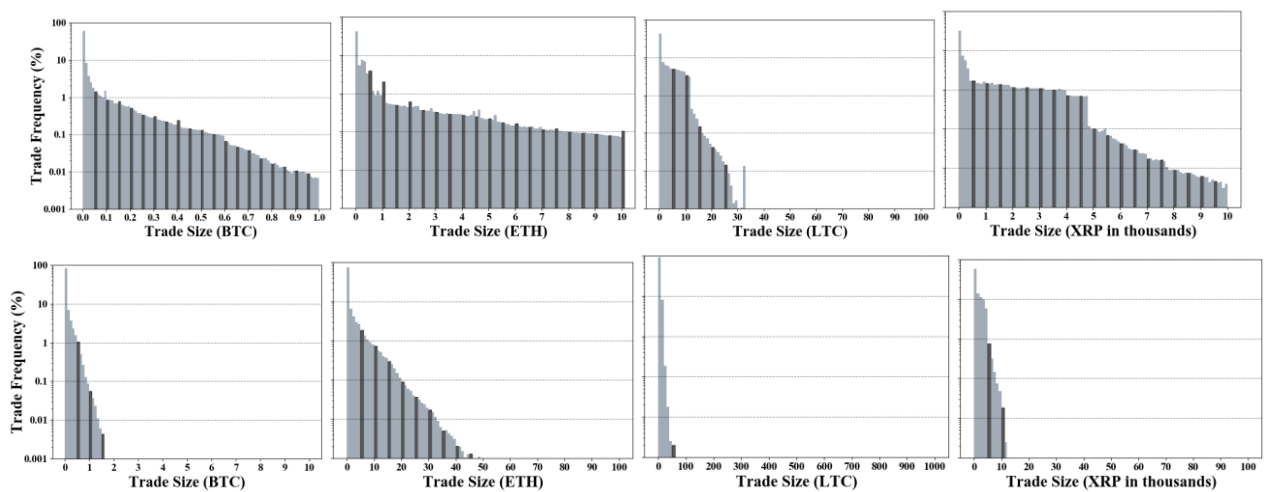

**Panel U: Unregulated Tier-2 Exchanges**
U8

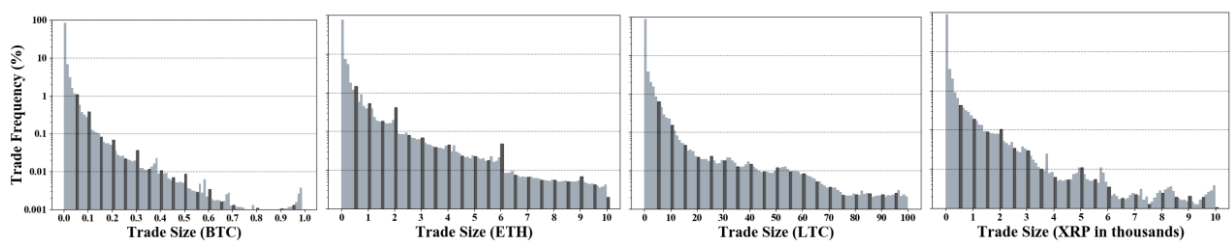

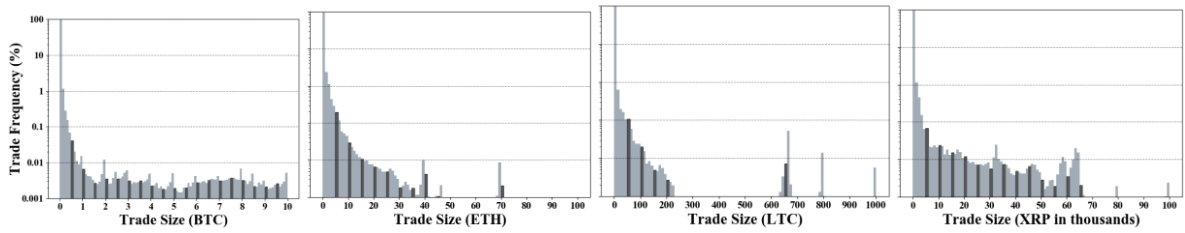

**U9**

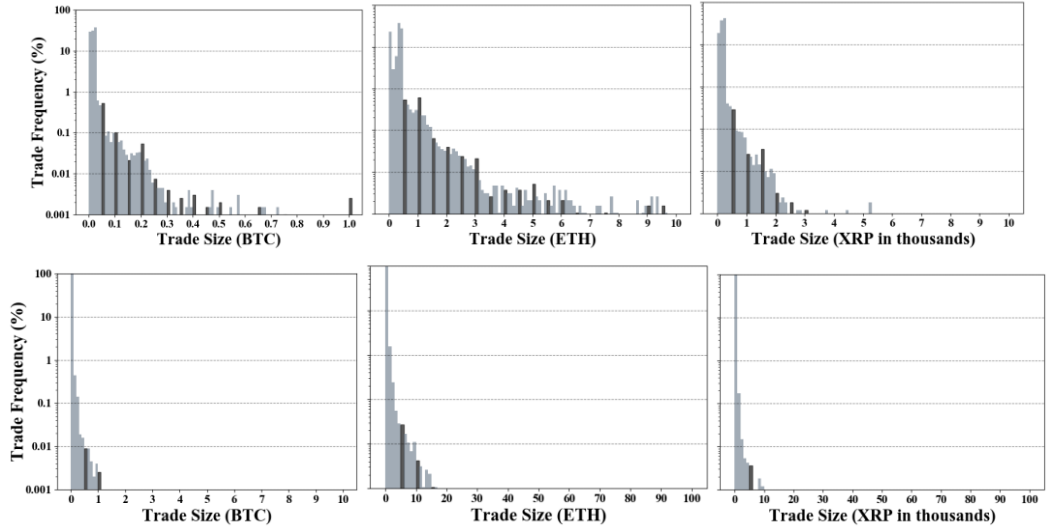

**U14**

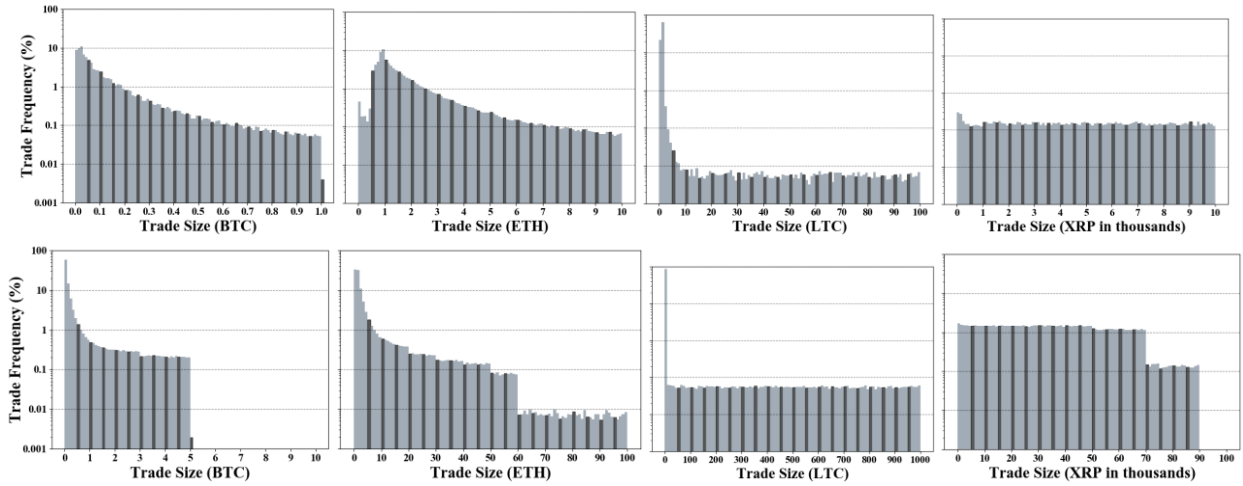

**Figure 3. Illustration of the *t*-test for Clusters**

Trade frequencies at round trade-sizes are tested against unrounded trade-sizes nearby. Frequency for trade-size *i* is calculated as the number of trades with size *i* over the total number of trades in an observation window (e.g. *i*-50 to *i*+50). Frequencies at round trade sizes (e.g. the 200[th] unit) and the highest frequencies of nearby unrounded trades (e.g. the 160[th] unit) are recorded as a pair. The *t*-test on the difference between round and unrounded frequencies in a pair is then carried out over a sample of all pairs.

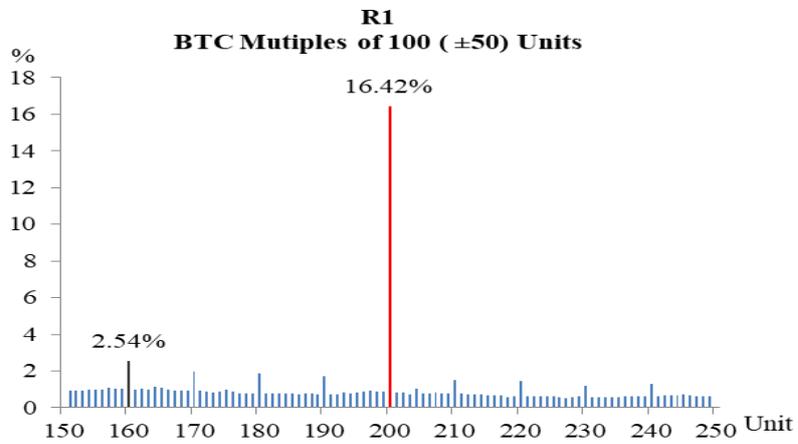

# Figure 4. Tail Distribution and Power-law Fitting

Figure 4 displays tails of trade-size distributions and the fitted power-law lines on log-log plots. Panel R, Panel UT, and Panel U show distribution of trade-size in regulated exchanges, Tier-1 unregulated, and Tier-2 unregulated exchanges, respectively. Regulated exchanges are those that are certified and regulated by the New York State Department of Financial Services. Unregulated exchanges are categorized into unregulated Tier-1 and unregulated Tier-2 exchanges based on website traffic ranks. For each crypto exchange, four trading pairs are presented, including BTC/USD, ETH/USD, LTC/USD, and XRP/USD. Fitted power-law lines are plotted with parameters estimated by Ordinary Least Square (OLS) and Maximum Likelihood Estimation (MLE), shown in black and red lines, respectively. Blue dots represent empirical data points for trade-size frequencies.

**Panel R: Regulated Exchange**

**R2**

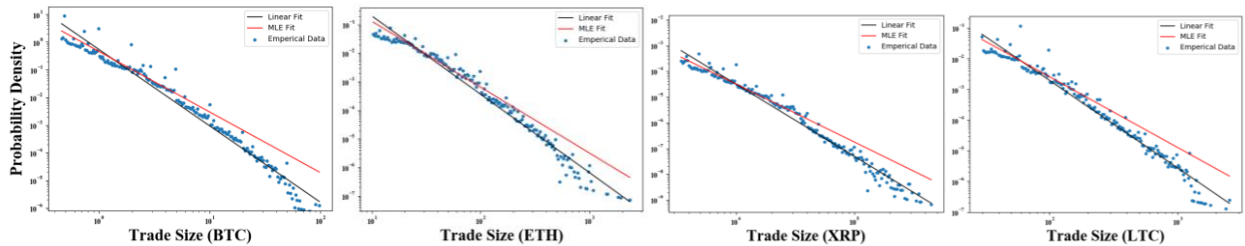

**Panel UT: Unregulated Tier-1 Exchanges**

**UT6**

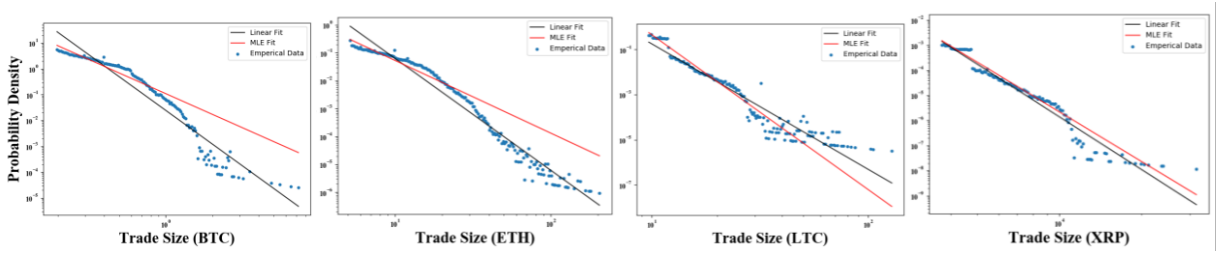

**Panel U: Unregulated Tier-2 Exchanges**

**U8**

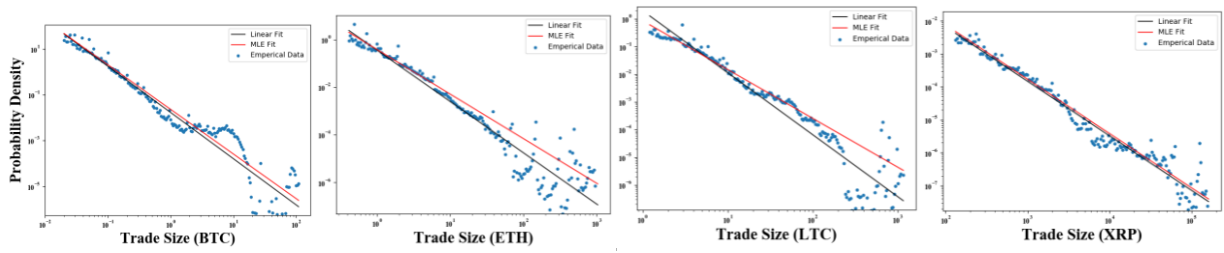

**U9**

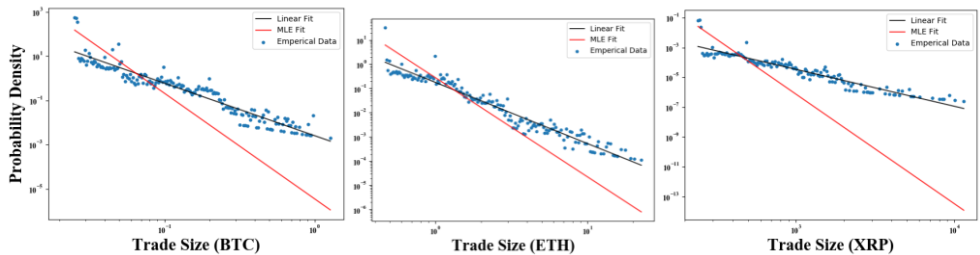

**U14**

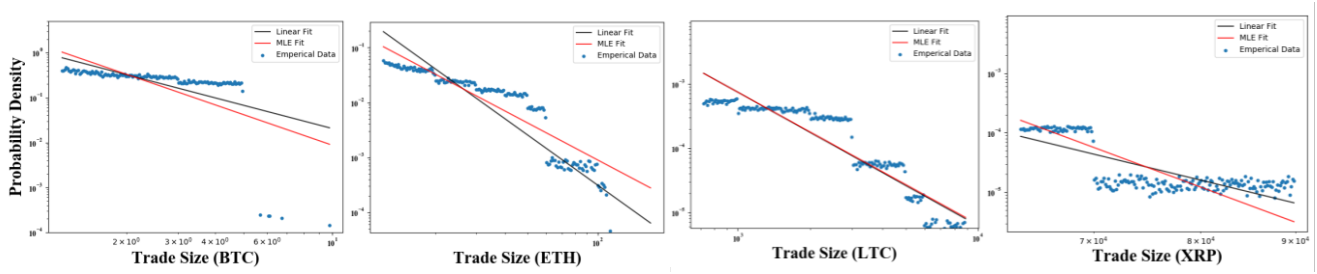

**Figure 5. Percentage of Failed Tests**

Figure 5 presents the percentage of failed tests for each crypto exchange. We summarize three statistical tests in Section 3 for each crypto exchange, including Chi-squared tests for Benford's Law, *t*-tests for trade size clustering, and scaling exponents for power-law fittings. For each test, we report four cryptocurrency pairs, BTC, ETH, LTC, and XRP. The test results are grouped by exchanges and cryptocurrencies, shown in two subplots. For each exchange (or cryptocurrency), the percentage of failed tests is calculated as the number of failed tests at a 5% significant level over the total number of tests. Specifically, in Chi-squared tests of first-significant-digits, 'failure' is when a distribution failed to conform to Benford's Law, statistically at the 5% significance level. In the *t*-test of clustering effect, 'failure' is when a distribution does not show apparent size clustering at multiples of 100 units at a 5% significance level. In power-law fitting tests, 'failure' refers to the situation when the scaling exponent ($\hat{\alpha}_{OLS}$ or $\hat{\alpha}_{Hill}$) is located outside the Pareto–Lévy range (1, 2).

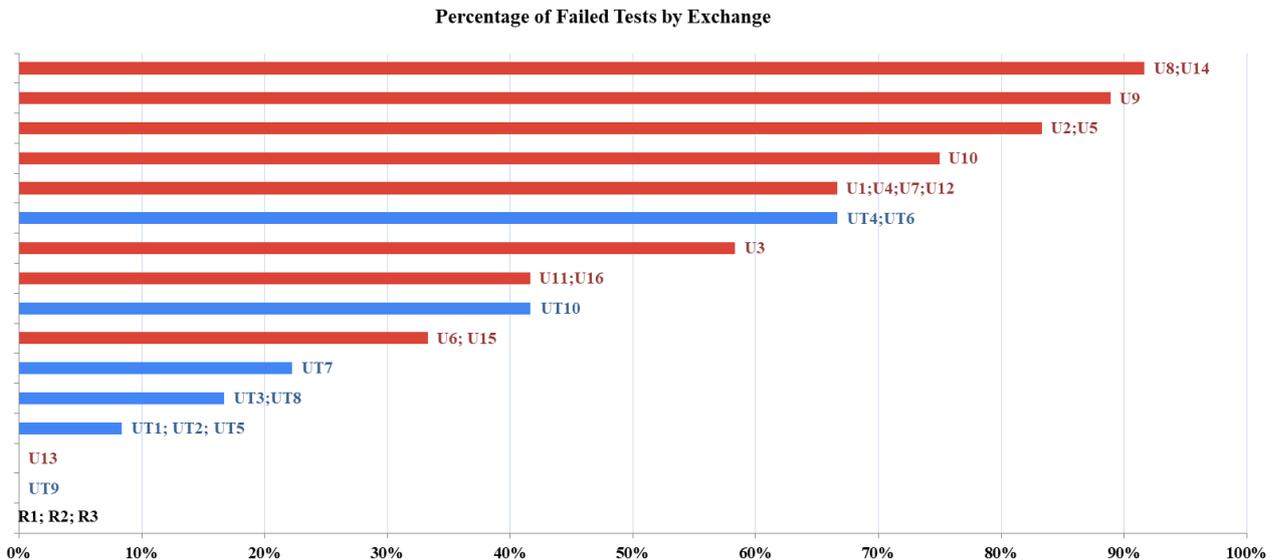

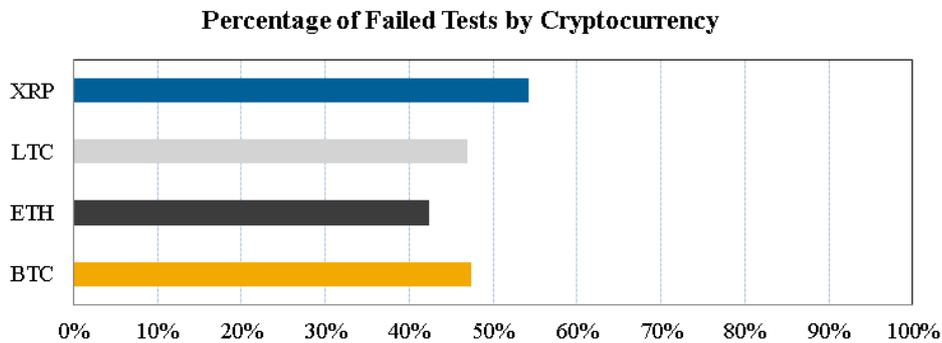

## Figure 6. Trading Volumes and Ranks

Figure 6 plots the quantitative relationship between (logarithm) trade volumes and exchange ranks. Data fitting is carried out with Ordinary Least Square (OLS) regression. The estimated coefficients are reported below (*t*-statistics in brackets) with an adjusted R$^2$ of 93%.

$$\text{Exchange rank}_i = 416.269 - 19.202 * \log(\text{Volume}_i) + \varepsilon_i$$

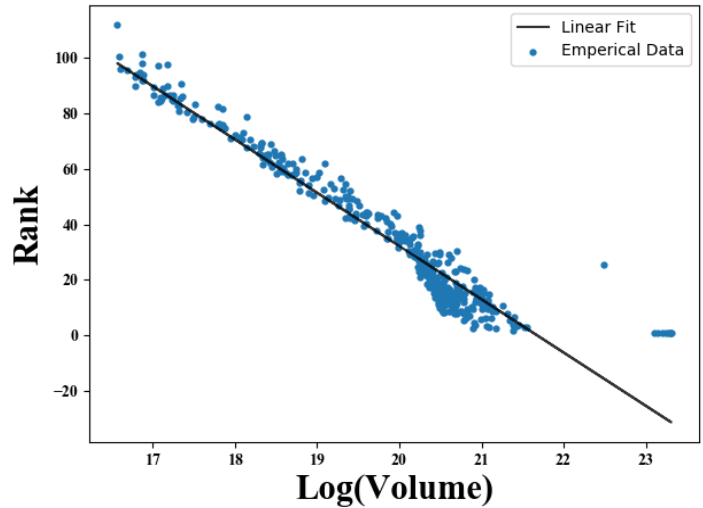

## Figure 7. Improvement in Ranks and Wash Trading

Figure 7 plots the relationship between the estimated fraction of wash trading and the improvement in counterfactual ranks. The counterfactual rank is estimated based on the estimated "real" volume, i.e., the difference between reported volume in CoinMarketCap and estimated wash trading volume, using the volume-rank relationship documented in Figure 6. Rank improvement is the difference between the counterfactual rank and reported rank in CoinMarketCap.

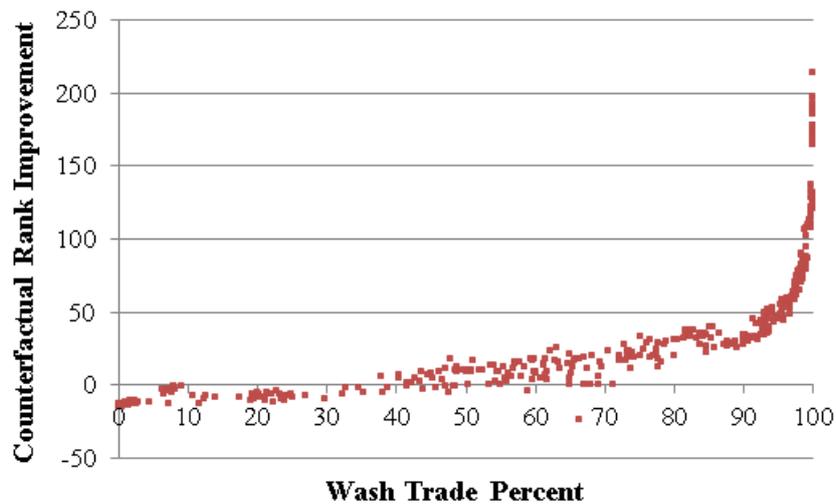

**Appendix A. Institutional Background of Crypto Exchanges: Development and Regulation**

We provide in this appendix the institutional background of crypto exchanges. Readers familiar with the cryptocurrency industry may skip reading.

Satoshi Nakamoto introduced Bitcoin in October 2008 and launched it three months later with one headline in the Times on January 3, 2009, "Chancellor on brink of second bailout for banks," embedded in the genesis block. Because Bitcoin is open-source, other "altcoins" (alternative to Bitcoin) quickly emerged to imitate or improve upon the first few cryptocurrencies. For example, Ethereum, EOS, and Tron were developed as public platforms for smart contracts and decentralized applications, with native cryptocurrencies on their own blockchains.[40] As we write, over 8000 cryptocurrencies have been launched and circulated globally. The total market capitalization of all cryptocurrencies just pasted $1 trillion in January 2021. Bitcoin alone once reached nearly $760 billion, larger than Visa ($452 billion on Jan 31, 2021) or Facebook ($736 billion on Jan 31, 2021).

The increasingly sophisticated crypto ecosystem is comprised of mining, payment companies, wallets, DApp (decentralized application), and crypto exchanges (Hileman and Rauchs, 2017), with increasing awareness and adoption among financial institutions and retail investors. Crypto exchanges — centralized gateways that facilitate money flow between fiat currency and (decentralized) cryptocurrency systems — play a critical and dominant role in the industry (Griffin and Shams, 2020). To date, over 300 exchanges provide cryptocurrency services around the globe, often with leverage facilities and derivatives on cryptocurrencies. Incumbents exit and new competitors keep emerging under loose regulatory standards. Because exchanges offer similar products and services, the competition is even fiercer than that in traditional markets.[41]

Currently, the total cryptocurrency trading volume on exchanges (likely in large part speculation activities) is much higher than the on-chain transaction volume (likely actual usage). With considerable traffic, exchanges usually hold a large number of various cryptocurrencies because of liquidity demand and custody for customers. Moreover, Initial Exchange Offerings (IEOs) have often substituted Initial Coin Offerings (ICOs) since 2019, in which an exchange may work with a start-up issuing cryptocurrencies or tokens.[42] As a result, they wield enormous power in the industry. This is somewhat ironic, given the initial ideals of decentralized trust and financial democratization.

---

[40] Monero, Zcash, and Dash were created to address Bitcoin's privacy limitations and shortcomings. Other cryptocurrencies focused on applications content creation and copyright (Steem, Ink), on social/communication (KEY, SNT), on the internet of things (IOTA, QTUM) and computation power/cloud storage (SC, FCT), among many others.

[41] Unlike established brands with user stickiness and network effect (Halaburda and Gandal, 2016; Cong, Miao, Tang, and Xie, 2019), newcomers (with little reputation) are more tempted to pursue high rankings that might be achieved via wash trading. Top ranked exchanges are thus not necessarily reputable and secure and investors who are misled to them could face substantial risks. For example, FCoin, which become insolvent in February 2020, previously ranked 56th on CoinMarketCap. However, Gemini, a crypto exchanged certified and regulated by the New York State Department of Finance, is listed 124th on the second page of CoinMarketCap. *https://coinmarketcap.com/rankings/exchanges/reported/2/* (Last accessed December 29, 2019)

[42] Security Token Offerings (STOs) in which token issuance is treated as a regular security issuance were hyped to be the new norm, but are limited by the heavy regulation. Initial DEX Offerings (IDOs, in which DEX stands for decentralized exchanges) have received attention since 2019 but are in limited scale and are not our focus.

Unregulated exchanges are not required to report trading records to any authority. However, due to business needs and peer competition, exchanges tend to be more transparent. For example, algorithmic trading needs high-frequency market data, which implies that exchanges need to feed data to traders through API portals. At the same time, market ranking websites and data aggregators such as the CoinMarketCap Data Accountability & Transparency Alliance are pushing exchanges for more transparency, accountability, and disclosure from projects.

In the early days, regulators deemed the cryptocurrency industry small and unimportant. It was widely believed that all crypto exchanges had, to some extent, engaged in non-compliant and unethical behavior (Gandal et al., 2018; Moore and Christin, 2013; Moore et al., 2018). Exchanges usually hold substantial funds from users' accounts (both in fiat and cryptocurrencies) without proper custody and insurance, which raises severe concerns. Moore and Christin (2013) and Moore et al. (2018) examine the failure of Bitcoin exchanges from 2010 to 2015 due to security breaches (including dominant exchanges such as Mt. Gox). Most often, implied counterparty risk manifests in the form of notorious 'runaway bosses' incidents or exit scams (malicious closure of exchanges and stealing users' funds). For example, the once largest transaction-mining exchange FCoin suddenly claimed insolvency with $130 million client's funds missing (Zhao, 2020).[43] Some exchanges get into legal quagmires through Ponzi schemes and scams. Xcoinx operated by the startup Onecoin is an example. Others include Coinroom (Alexandre, 2019), Cobinhood (Palmer, 2020), OKUEX, and Soxex. The list goes on.

Profit-driven exchanges may also take advantage of the information asymmetry or even directly act against users' interests through various market manipulation measures. In an unregulated environment, an unethical cryptocurrency exchange can be "both a referee and a player" at the same time. Gandal et al. (2018) investigate the manipulative trading in Mt. Gox, a Bitcoin exchange, over the period from February to November 2013, and find that a suspicious trader called "Markus," most likely an exchange-owned account, participated in manipulative trading. Our paper also shows that many exchanges have engaged in wash trading, likely aiming to improve their ranking or to attract more customers.

How do exchanges wash trade? The most primitive and rough approach is to simply print trading records (which do not really happen) in the trading history data. This approach was easily discovered by customers and observers to monitor live trade books from exchanges' websites. Even if exchanges put fake orders into the order book and later fill these orders themselves, such a practice is limited to approved accounts (exchange owned) that can fill these orders. This approach can be detected based on the mismatching between order book depth and trade spread. For example, some industrial reports utilize the relationship between exchange trading volume and liquidity (spread) for detecting wash trading. A more technically involved way of wash trading is to deploy the algorithm trading robot to create real orders and execute wash trades on diverse accounts. Exchanges can deploy wash-only robots or insert wash trades into their market-making robots every now and then. However, this approach entails the risk of loss if the positions are not closed in time. Finally, as mentioned earlier, some exchanges provide incentives for their users to (wash) trade by various fee rebates or transaction-mining programs. A combination of the above actions makes it extremely hard to detect specific wash trades with transaction history alone.

---

[43] Transaction-mining is when an exchange provides incentives to users, usually in the format of exchange issued token.
There are debates on transaction mining, ethically and financially. It is an original scheme from cryptocurrency exchanges that combines token distribution, dividend distribution and user incentives. It can help newly established exchanges to bootstrap the operation and obtain clients fast. However, without proper regulation, it inevitably lead to wash trading.
Some transaction mining exchanges deliberately make the reward override trading fees. As a result, a large portion of users trade for the sole purpose of getting transaction mining reward. The most famous transaction mining exchange Fcoin get $5.6 billion daily trading volume in less than a month from its establish, that is more than the sum of the rest top-10 platforms on CoinMarketCap. (https://www.coindesk.com/new-crypto-exchange-draws-fire-over-controversial-business-model)

The general lack of consumer protection in the cryptocurrency industry aggravates the situation. Consumers' legitimate rights and interests heavily rely on exchanges' self-discipline and good faith. If user interests are undermined in incidents such as hacking or bankruptcy, victims get little compensation from either exchanges or third-party insurance companies.

As such, risks in the cryptocurrency exchange ecosystem have drawn significant attention from regulatory authorities in recent years. Regulators in multiple jurisdictions have published statements to warn the public about the risks (Yu, 2018), and have built internal divisions and created new institutions to closely monitor the development of the cryptocurrency industry (Brett, 2019). Authorities (e.g., Bank of Canada, UK Financial Conduct Authority, New York Federal Reserve Bank) have conducted surveys to investigate the awareness and adoption of cryptocurrency among retail and institutional investors. In a July 2018 report to the G20, Mark Carney, the chair of the Financial Stability Board and the head of the Bank of England, warned that illegal manipulations in equity markets are rampant in crypto: wash trading, pump and dumps, and spoofing by traders (mostly bots) are particularly detrimental to financial stability and robustness to crises and recessions (Rodgers, 2019). Since 2017, official cryptocurrency documentation and guidelines have been released by regulatory agencies in around 20 countries and territories, including the United States, European Union, United Kingdom, China, Japan, etc. (Blandin et al., 2019).

Wash trading could be a major challenge for regulators because the unique features of the cryptocurrency industry render traditional attempts futile and ineffective. [44] For one, regulatory frameworks are different across countries without a consensus on the correct approach. The intention and infrastructure for sharing information and collaborative effort are also lacking among regulators in different countries.

Industry leaders also took action to fight the wash trading problem. CoinMarketCap, for example, introduced a mandatory API program for all listed exchanges to improve credibility and transparency (CMC, 2019a). They later developed another rank algorithm based on exchanges' liquidity instead of volume (CMC, 2019b). CryptoCompare, a British cryptocurrency data analysis firm, launched a unique exchange benchmark product that would help safeguard against false exchange volume reports (Tsavliris, 2019). Nomics, a data provider, developed Transparency Volume based on their ranking criteria, claiming it is less likely to include wash trading volume (Nomics, 2019). Nonetheless, the industry is in dire need of effective regulatory tools and a well-integrated regulatory framework.

---

[44] The United States banned wash trading in the Commodity Exchange Act (CEA) 1936, and the European Union listed it in the Market Abuse Directive No 2003/6/EC, etc. Therefore, financial services that are operating under the traditional regulatory framework are naturally prohibited from wash trading.

# Appendix B. Exchange Names

The table here displays exchange codes and the corresponding name of crypto exchanges in the data set.

| Exchange Code | Exchange Name |
|---|---|
| **Panel A Regulated exchanges** | |
| R1 | Bitstamp |
| R2 | Coinbase |
| R3 | Gemini |
| **Panel B Unregulated Tier-1 exchanges** | |
| UT1 | Binance |
| UT2 | Bittrex |
| UT3 | Bitfinex |
| UT4 | HitBTC |
| UT5 | Huobi |
| UT6 | KuCoin |
| UT7 | Liquid |
| UT8 | Okex |
| UT9 | Poloniex |
| UT10 | Zb |
| **Panel C Unregulated Tier-2 exchanges** | |
| U1 | Bgogo |
| U2 | Biki |
| U3 | Bitz |
| U4 | Coinbene |
| U5 | DragonEX |
| U6 | Lbank |
| U7 | Mxc |
| U8 | Fcoin |
| U9 | Exmo |
| U10 | Coinmex |
| U11 | Bibox |
| U12 | Bitmart |
| U13 | Bitmax |
| U14 | Coinegg |
| U15 | Digifinex |
| U16 | Gateio |